\documentclass[APA,Times1COL]{WileyNJDv5} 
\usepackage{changes}
\usepackage{amsmath}
\usepackage{booktabs}
\usepackage{longtable}
\usepackage[authoryear,longnamesfirst]{natbib}
\usepackage{verbatim}
\newcommand{\new}[1]{{\color{black}#1}}

\articletype{Original Article}%

\received{xx xx xx}
\revised{xx xx xx}
\accepted{xx xx xx}
\journal{Journal}
\volume{00}
\copyyear{2025}
\startpage{1}

\begin{document}

\title{Emergent Learner Agency in Implicit Human–AI Collaboration: How Supportive and Contrarian AI Personas Reshape Interaction}

\author[1]{Yueqiao Jin}
\author[1]{Roberto Martinez-Maldonado}
\author[2]{Dragan Gašević}
\author[3]{Xibin Han}
\author[3,1]{Lixiang Yan}

\authormark{Jin \textsc{et al}}

\address[1]{\orgdiv{Faculty of Information Technology}, \orgname{Monash University}, \orgaddress{\city{Melbourne}, \country{Australia}}}
\address[2]{\orgdiv{Faculty of Education and School of Computing \& Data Science}, \orgname{The University of Hong Kong}, \orgaddress{\city{Hong Kong}, \country{China}}}
\address[3]{\orgdiv{School of Education}, \orgname{Tsinghua University}, \orgaddress{\city{Beijing}, \country{China}}}

\corres{Corresponding author Lixiang Yan, \orgdiv{School of Education}, \orgname{Tsinghua University}, \orgaddress{\state{Beijing}, \country{China}} \email{lixiangyan@tsinghua.edu.cn}}

\titlemark{Emergent Learner Agency in Implicit Human–AI Collaboration}

\abstract[Abstract]{\textbf{Background}

As agentic artificial intelligence (AI) systems move beyond tool-like support toward more autonomous, team-like roles, a central question concerns the extent to which such systems can meaningfully participate in collaborative learning. Emerging work suggests that agentic AI may adopt distinct interactional personas, such as supportive or contrarian roles, yet little is known about how these personas shape learner agency and group dynamics when AI operates as an undisclosed teammate.

\textbf{Objectives}

This study investigates how supportive and contrarian AI personas influence emergent learner agency, discourse patterns, and experiential outcomes in implicit human–AI creative collaboration. 

\textbf{Methods}

A total of 224 university students were randomly assigned to 97 online triads in human-only, supportive-AI, or contrarian-AI conditions. Teams completed an individual–group–individual creative movie-plot task via a 10-minute text chat. Discourse was coded using a creative–regulatory framework and analysed via transition network analysis, sequential pattern mining, and Gaussian mixture clustering to characterise emergent agency patterns. These patterns were linked to cognitive load, psychological safety, teamwork satisfaction, and creative performance.

\textbf{Results and Conclusions}

The findings revealed that contrarian AI produced challenge- and reflection-rich discourse indicative of productive friction, whereas supportive AI promoted agreement-centred trajectories. However, while contrarian personas stimulated critical engagement, they reduced teamwork satisfaction and psychological safety without yielding corresponding gains in creative performance. Educational designs need to balance epistemic challenge with the preservation of the affective climate to ensure that hybrid collaboration supports rather than undermines the learner experience.
}


\maketitle

\bmsection*{Lay Summary}

\noindent\textbf{What is currently known about this topic:}
\begin{itemize}
    \item Generative AI can participate as a peer-like contributor in collaborative learning, shaping how groups generate, evaluate, and integrate ideas.
    \item Supportive and challenging (contrarian) facilitation approaches are both useful in collaborative creativity, but they tend to trade off relational comfort versus critical engagement.
    \item Most evidence comes from explicit AI use; when learners know an AI is present, expectations and strategy shifts can confound observed collaboration processes.
\end{itemize}

\noindent\textbf{What this paper adds:}
\begin{itemize}
    \item When AI participation is implicit (not disclosed), AI personas still reorganise collaboration: contrarian AI pulls discourse into challenge- and reflection-linked pathways, while supportive AI stabilises agreement-centred trajectories.
    \item Challenge-oriented discourse was predominantly produced by AI agents, whereas reflective regulation (meta-level monitoring) remains uniquely human in this setting.
    \item Contrarian AI reduces teamwork satisfaction and psychological safety, without corresponding gains in cognitive load reduction or creative performance.
\end{itemize}

\noindent\textbf{Implications for practice or policy:}
\begin{itemize}
    \item Treat AI personas as governance knobs for group discourse: use supportive personas to maintain cohesion and momentum, and use contrarian challenge sparingly or later in the task when norms and trust are established.
    \item If challenge is desired, implement bounded friction: constrain frequency/intensity of critique, pair challenges with integrative prompts, and add repair moves (e.g., acknowledgement, summarising, option-generation) to protect psychological safety.
    \item When AI may influence collaboration invisibly (e.g., writing assistants, recommender systems), build learners’ meta-collaborative literacy (how to interpret, accept/reject, and retain ownership of suggestions) and provide transparency/consent options where feasible.
\end{itemize}

\section{Introduction}
Collaborative learning is a complex dynamic system characterised by the multifaceted interplay of cognitive, social, and emotional processes \citep{kaplan2020steps, hilpert2018complex}. As the digital landscape undergoes a vast transformation marked by the integration of Artificial Intelligence (AI), the environments in which these dynamics unfold are fundamentally changing \citep{chen_two_2022, ouyang_artificial_2021}. In higher education and professional settings, AI systems are shifting from passive tools to active participants embedded within collaborative processes \citep{kamalov_evolution_2025, sapkota_ai_2025, xi_rise_2025}. This shift is particularly visible in emerging work on agentic AI, where systems can simulate human-like contributions to support essential 21st-century skills developments, such as creative problem solving and teamwork, that have traditionally relied on human-human interaction \citep{park_generative_2023, wei_effects_2025, laal_21st_2012}. While this agentic capability offers new opportunities to actively shape interactional processes, for instance, by steering groups through necessary phases of divergent thinking and convergent synthesis \citep{farrokhnia_improving_2025}, it introduces a new frontier: implicit or awareness-free AI participation, where algorithms contribute to group work without revealing their artificial identity \citep{hwang_ideabot_2021, schecter_how_2025, zhang_human_2023}.

Investigating awareness-free AI participation in collaborative learning is critical for both scientific and practical reasons. Scientifically, implicit participation allows for the study of fundamental mechanisms of social influence and epistemic contribution uncontaminated by ``expectancy effects'', the preconceptions learners hold about AI competence or authority \citep{rubin_comparing_2025, jakesch_human_2023, zhai_effects_2024}. Practically, it offers a powerful experimental lens into a future where AI is ubiquitously woven into the fabric of digital communication platforms \citep{chen_two_2022, ouyang_artificial_2021}. In these emerging environments, AI can increasingly shape learner contributions, via auto-complete, content moderation, or peer-matching, without announcing its presence \citep{hwang_ideabot_2021, schecter_how_2025, zhang_human_2023}. This raises an urgent question for educational practice: how does AI redirect or redistribute learner agency when students cannot explicitly regulate or negotiate the AI's involvement?

In collaborative learning, learner agency is not a static individual trait but an emergent, interactional phenomenon arising from the feedback loops between participants \citep{darvishi_impact_2024, jarvela_human_2023}. It is enacted through the complex interplay of divergent processes (generating ideas, exploring alternatives) and convergent processes (evaluating contributions, integrating perspectives) \citep{farrokhnia_improving_2025, korde_alternating_2017}. Successful collaboration relies on the system's ability to self-organise between these states: moving from social co-regulation that builds safety to productive friction that deepens reasoning \citep{ward_productive_2011, holtz_using_2018}. However, these dynamics have historically been studied in human-only systems \citep{dillenbourg_what_1999, roschelle_construction_1995}. When the ``peer'' in the system is an AI enacting a specific persona, such as a supportive facilitator or a contrarian challenger, the feedback loops that sustain agency may be fundamentally altered \citep{joo_ai_2025, hwang_ideabot_2021, weijers_intuition_2025}. By introducing stable behavioural tendencies, AI agents may redistribute epistemic labour and reshape the temporal evolution of discourse in ways that are difficult to capture through traditional pre-post comparisons alone \citep{yang_ink_2024, schecter_how_2025}.

Despite the rapid proliferation of human-AI collaboration research, a significant theoretical and methodological gap remains. Most studies focus on explicit AI assistance and outcome-oriented metrics, often neglecting the \textit{temporal dynamics} and \textit{processual nature} of the interaction \citep{yang_ink_2024, yan_distinguishing_2025, molenaar_towards_2022}. The existing literature lacks a psychologically grounded understanding of how AI personas, operating as invisible distinct social-epistemic forces, influence the emergent organisation of group agency \citep{joo_ai_2025, darvishi_impact_2024, yan_beyond_2025}. Does a supportive AI stabilise the system into a ``safe'' but stagnant equilibrium \citep{bai_enhancing_2024, jarvela_human_2023}? Does a contrarian AI introduce necessary non-linear perturbations that trigger deeper synthesis, or does it disrupt the emotional cohesion required for agency to flourish \citep{ward_productive_2011, holtz_using_2018, weijers_intuition_2025}? Addressing these questions requires analytic approaches that foreground temporality and interactional process, examining how collaboration unfolds over time, how earlier contributions shape later possibilities, and how local discourse moves accumulate into system-level patterns of agency. \citep{cukurova_interplay_2025}.

To address this lag in theoretical and methodological integration, this study examines the dynamics of learner agency in implicit human-AI creative collaboration through the lens of complex dynamic systems that treats agency as an emergent property of temporally evolving interaction rather than a static individual attribute. Methodologically, we employ a multi-method approach combining Transition Network Analysis (TNA) to model the probabilistic pathways of regulatory states, and theory-driven Sequential Pattern Mining to trace the evolution of agency over time. By investigating how supportive and contrarian AI personas reconfigure the structural, temporal, and profile-level manifestations of agency, and linking these process dynamics to cognitive, affective, and creative outcomes, we aim to bridge the gap between novel data-driven insights and foundational educational psychology. 

\section{Background} 

\subsection{Hybrid Human-AI Collaboration and Implicit AI Participation}
\label{sec:implicit}
The increasing sophistication of generative AI has accelerated interest in hybrid human-AI collaboration across education, industry, and creative domains \citep{yan_promises_2024, molenaar_towards_2022}. In many knowledge-intensive settings, AI systems are no longer peripheral tools but active contributors that participate in ideation, discussion, and decision-making processes \citep{wang_impact_2025, kim_enhancing_2024}. Early work in human-AI collaboration has shown that AI can enhance the breadth of ideas generated, scaffold group discussion, and support more systematic exploration of problem spaces \citep{farrokhnia_improving_2025, nguyen_human-ai_2024}. In educational contexts, AI has been integrated as a peer-like discussion partner, facilitator, or challenger, often with promising effects on engagement and productivity \citep{joo_ai_2025, weijers_intuition_2025, bai_enhancing_2024}. These developments mirror broader trends in agentic AI research, where autonomous or semi-autonomous agents are used to simulate social interactions, classroom environments, and civic behaviours to better understand how artificial and human actors jointly shape learning and decision-making \citep{park_generative_2023, xi_rise_2025, sapkota_ai_2025}.

Despite this progress, most empirical studies assume explicit AI use: participants knowingly interact with an AI system, bringing with them preconceptions about its competence, authority, or limitations \citep{jakesch_human_2023, zhang_human_2023}. As a result, learner behaviours may reflect strategic adaptation to the known presence of AI rather than their default interactional patterns \citep{rubin_comparing_2025, schecter_how_2025}. In contrast, implicit or awareness-free AI participation hides the agent’s identity, allowing researchers to examine foundational mechanisms of influence, coordination, and social meaning-making that may be masked in explicit-use contexts \citep{hwang_ideabot_2021, schecter_how_2025, zhang_human_2023}. Such implicit setups are increasingly relevant to real-world educational environments. AI-driven recommendation engines, automated feedback systems, and peer-learning platforms may shape learners’ contributions without announcing their involvement, subtly influencing discourse, decision-making, or perceptions of group dynamics \citep{chen_two_2022, ouyang_artificial_2021}. Understanding these implicit effects is essential for anticipating how AI may redistribute agency, authority, or responsibility within collaborative learning \citep{darvishi_impact_2024, yan_beyond_2025}.

Empirically, however, we know little about how AI presence, particularly persona-driven AI behaviour, can reshape the interactional organisation of small-group collaboration when learners cannot identify the source of contributions. Existing work has documented changes in idea diversity or task outcomes but rarely examines the deeper social-epistemic processes through which collaboration unfolds \citep{tian2025impact}. This gap motivates a closer analysis of how implicit AI participation transforms the fabric of creative discourse, especially during activities that traditionally rely on human-human synergy, such as story generation or open-ended problem solving.

\subsection{Collaborative Creativity and Regulation}

\label{sec:creativity}

Collaborative creativity places distinctive demands on group interaction because progress depends not only on generating novel ideas but also on regulating how those ideas are evaluated, combined, and advanced toward a shared outcome \citep{barrett2021creative, craft2008studying}. Unlike routine problem solving, creative collaboration involves sustained uncertainty, multiple plausible directions, and evolving criteria of quality \citep{wiltschnig2013collaborative, tang2020developing}. Groups must therefore manage both epistemic work (what ideas are proposed and how they develop) and process-level coordination (when to explore, when to consolidate, and how to maintain collective momentum) \citep{barrett2021creative, hilliges2007designing}. Prior research highlights that breakdowns in creative collaboration often stem less from a lack of ideas than from failures to regulate transitions between exploration, evaluation, and synthesis, leading to premature consensus, unresolved disagreement, or stalled progress \citep{hilliges2007designing, vass2007exploring}. This makes creative activity a particularly revealing context for examining how interactional regulation shapes collective sense-making.

From a process-oriented perspective, regulation in collaborative creativity is enacted through functional patterns of discourse rather than through explicit role assignment or formal planning \citep{guan2024creating, volet2009high, jarvela2023predicting}. Interaction unfolds as sequences of conversational moves that open the idea space, negotiate alignment, introduce critique, consolidate alternatives, and occasionally step back to monitor progress or recalibrate strategy \citep{barrett2021creative, sawyer2023explaining}. The relative availability and ordering of these functions determines whether groups sustain productive tension between divergence and convergence or drift toward either uncritical agreement or fragmented exploration \citep{de2019scientific, kopcso2017regulated}. Importantly, these regulatory patterns are sensitive to the interactional environment: small shifts in participation norms or response tendencies can systematically bias groups toward particular trajectories \citep{guan2024creating, barron_when_2003, graesser_advancing_2018}. This sensitivity makes collaborative creativity an analytically powerful setting for studying how external contributors, such as AI systems embedded in group interaction, may subtly reweight regulatory pathways, shaping not only what is discussed but how collaboration unfolds over time.

\subsection{Learner Agency in Collaborative Learning}
\label{sec:agency}

Learner agency is widely recognised as a central driver of productive collaborative learning \citep{darvishi_impact_2024, yan_beyond_2025}. Agency encompasses learners' capacity to initiate ideas, influence the direction of group work, evaluate and refine contributions, and regulate collective progress \citep{jarvela_human_2023, molenaar_concept_2022}. Importantly, agency is enacted through discourse: utterances that propose, question, elaborate, integrate, or reflect constitute the micro-level behaviours through which influence and ownership unfold \citep{weinberger_framework_2006, baker_argumentative_2009}. Work in collaborative learning and creativity further conceptualises learner agency emerges through the coordinated interplay of divergent processes that expand the idea space and convergent processes that evaluate, prioritise, and synthesise contributions into coherent shared outcomes \citep{hansen2022students, farrokhnia_improving_2025}.

From this perspective, agency is not a static individual attribute but a dynamic, interactionally constituted process that unfolds through cycles of divergence and convergence. The literature consistently shows that productive collaboration requires a dynamic interplay between divergent and convergent processes, \citep{roschelle_construction_1995, dillenbourg_what_1999}. Divergence without convergence leads to unfocused ideation, whereas convergence without divergence suppresses originality. Creative tasks, such as joint storytelling or brainstorming, exemplify this delicate balance \citep{yang_ink_2024, wei_effects_2025}. Prior research identifies distinct discourse moves that support these processes, including idea generation, elaboration, challenge, agreement, integration, and reflection \citep{weinberger_framework_2006, noroozi_facilitating_2013}. These moves collectively reveal how agency is distributed among participants, who contributes epistemically, who shapes directionality, and who supports or regulates group progress \citep{barron_when_2003}.

Empirical studies demonstrate that agency fluctuates over time, emerging from the moment-to-moment structure of interaction rather than from stable individual characteristics \citep{jeong_seven_2016, chan_peer_2001}. Groups may gravitate toward specific patterns: some dominated by supportive elaboration, others characterised by frequent challenge and synthesis \citep{ward_productive_2011, holtz_using_2018}. These patterns influence the quality of creative outcomes, the degree of shared understanding, and the emotional climate of collaboration. Yet, almost all empirical investigations of such processes rely on human-only teams. When an AI system participates, particularly one that consistently enacts a persona, how the distribution and enactment of agency shifts remains unclear \citep{joo_ai_2025, hwang_ideabot_2021, brandl_can_2025}. For instance, supportive AI may dominate affiliative moves, subtly dampening human contributions, while contrarian AI may provoke conflict that humans feel less comfortable responding to \citep{weijers_intuition_2025}. Understanding how these interactions unfold is crucial for determining whether AI enhances or erodes learners' opportunities to exercise agency. This lack of empirical evidence on how AI-driven behaviours manifest in creative-regulatory discourse, and how they shape the emergent roles of human participants, underscores the need for systematic analysis of implicit hybrid collaboration.

Recent research on AI-assisted collaborative learning further suggests that generative AI systems actively reshape how learner agency is enacted in group contexts, not only by contributing content but by reconfiguring interaction patterns and epistemic participation. Studies of AI-supported collaborative writing and problem-solving show that generative agents can increase cognitive engagement, scaffold reflective thinking, and support shared metacognition, thereby altering the distribution of epistemic and regulatory contributions within groups \citep{Hu2024Enhancing, Lin2024Enhancing, Iqbal2025Generative}. At the same time, AI participation can shift interactional dynamics by introducing new conversational roles, such as co-writer, facilitator, or simulated peer, which influence how learners initiate, respond to, and build upon ideas \citep{Kim2024Exploring, Kharrufa2024LLMs, Shen2025The}. However, these benefits are accompanied by emerging concerns: overreliance on AI may reduce peer-to-peer interaction or critical engagement, and the presence of AI-generated contributions can subtly redistribute ownership and agency within collaborative processes \citep{Zhai2024The, Pham2025The, Giannakos2024The}. Taken together, this emerging body of work highlights that learner agency in AI-mediated collaboration is not merely supported but fundamentally reorganised by the affordances and behavioural patterns of generative systems, underscoring the need to examine how such influences unfold at the level of discourse and interaction over time.

To avoid conceptual overreach, this study operationalises learner agency in a deliberately bounded way: not as a total account of human agency, but as its behavioural manifestation in collaborative discourse. Specifically, we focus on how agency becomes visible through observable discourse moves by which participants introduce, challenge, elaborate, integrate, and regulate ideas in joint activity. This means that our analysis does not attempt to directly capture internal planning, unspoken reflection, identity development, or longer-term intentionality, which are important aspects of agency in broader theoretical accounts but are not directly measurable from interaction traces alone. Instead, consistent with sociocultural and dialogic perspectives, we conceptualise learner agency as an interactional and relational accomplishment enacted within activity systems and negotiated through moment-to-moment positioning in talk, where influence, ownership, and responsibility are continuously distributed and re-distributed across collaborators \citep{vygotsky_mind_1978, tomasello_becoming_2019, kelly_what_2006}. Accordingly, our coding framework (Section~\ref{sec-code}) uses creative-regulatory discourse moves as process indicators of two complementary dimensions of emergent agency: epistemic agency, reflected in contributions that shape the conceptual direction of the shared task (e.g., initiating, challenging, integrating ideas), and regulatory agency, reflected in contributions that monitor, coordinate, or redirect how the group proceeds (e.g., reflective monitoring and meta-level coordination) \citep{weinberger_framework_2006, molenaar_concept_2022}. This operationalisation is therefore intentionally partial, capturing behavioural manifestations of agency in interaction without directly measuring learners' intentions, identity-based meanings, or longer-term autonomy beyond the focal task, but is analytically appropriate for examining how agency is distributed and reconfigured in implicit human-AI collaboration.

\subsection{AI Personas as Social-Epistemic Forces in Collaboration}
\label{sec:personas}
AI personas, configurations that encode stable behavioural tendencies such as being supportive or contrarian, provide a powerful lens for understanding how AI shapes group interaction \citep{joo_ai_2025, shanahan_role_2023}. Personas influence not only linguistic style but also epistemic stance, pace of contribution, and patterns of response \citep{hwang_ideabot_2021, salvi_conversational_2025}. In human collaboration research, supportive behaviours such as affirmation, positive feedback, and elaborative building are known to enhance cohesion, trust, and coordination \citep{weinberger_framework_2006, baker_argumentative_2009}. Conversely, contrarian behaviours, questioning, challenging, introducing alternative perspectives, often stimulate deeper reasoning, perspective-taking, and integrative thinking \citep{ward_productive_2011, holtz_using_2018}. Both roles play essential functions in productive group work \citep{dillenbourg_what_1999, roschelle_construction_1995}.

However, when these behaviours are enacted by AI, the implications can differ in important ways. Humans may attribute undue authority or consistency to AI-generated statements, amplify or suppress their own contributions, or interpret AI behaviours through expectations shaped by technological interaction rather than peer interaction \citep{jakesch_human_2023, rubin_comparing_2025, zhang_human_2023}. AI agents, unconstrained by social fatigue or emotional vulnerability, may also enact their personas more rigidly than humans would, exerting disproportionate influence over group discourse \citep{schecter_how_2025, haupt_consumer_2025}. In implicit contexts, the effect may be even stronger: participants who believe they are collaborating exclusively with humans cannot modulate their responses based on known AI affordances, such as reduced sensitivity or increased resilience to disagreement \citep{hwang_ideabot_2021, schecter_how_2025, zhang_human_2023}.

Emerging empirical work suggests that AI behaviour can alter group norms, direct attention, and influence perceptions of task difficulty or partner competence \citep{weijers_intuition_2025, brandl_can_2025}. Yet, systematic evidence linking AI personas to emergent patterns of agency, structural, temporal, and role-based, remains sparse \citep{yan_beyond_2025, darvishi_impact_2024}. We lack robust understanding of whether persona-driven AI contributions give rise to distinctive agency profiles--recurring patterns in how individuals participate, influence direction-setting, and regulate collaborative processes over time--and how these profiles align with or diverge from human behaviour in creative collaboration \citep{Heinim_ki_2021, yang_ink_2024, wei_effects_2025}. These gaps motivate a deeper analysis of how AI personas function as social-epistemic forces that reshape agency distributions in hybrid teams.

\subsection{Consequences for Learner Experience and Educational Design}
\label{sec:consequence}

Understanding how AI reshapes collaborative dynamics is not only theoretically important but also crucial for educational design and policy. Collaborative creativity is a foundational 21st-century skill \citep{trilling_21st_2009, council_education_2012}, and AI-enabled environments increasingly mediate when, how, and with whom students collaborate \citep{molenaar_towards_2022, cukurova_interplay_2025}. If AI personas subtly steer discussions, introduce conflict, or dominate integrative moves, these effects may shape learners' experiences in ways that go unnoticed yet carry significant implications for participation equity, self-efficacy, and long-term comfort with collaborative work \citep{darvishi_impact_2024, yan_beyond_2025}.

Affective dimensions, including teamwork satisfaction and psychological safety, are particularly vulnerable to shifts in interactional climate \citep{barron_when_2003, graesser_advancing_2018}. Supportive personas may create smoother, more harmonious collaboration but risk fostering complacency or reducing critical engagement \citep{ward_productive_2011, holtz_using_2018}. Contrarian personas may deepen reasoning but simultaneously undermine comfort, trust, or perceived belonging \citep{weijers_intuition_2025, brandl_can_2025}. Cognitive load may also vary depending on whether AI contributions streamline or complicate group reasoning \citep{sweller_element_2010, stadler_cognitive_2024}. And while creative performance is a valued outcome, little is known about whether AI-enhanced discourse structures translate into measurable creative gains for learners \citep{wei_effects_2025, farrokhnia_improving_2025}.

\subsection{Research Questions}
Taken together, the preceding sections highlight a series of unresolved questions about how agentic AI, particularly when operating implicitly and enacting stable personas, reshapes collaborative processes, learner agency, and learner experience. While prior work has examined outcomes of human–AI collaboration, there remains limited understanding of how AI personas reconfigure the internal organisation of discourse, how these effects unfold temporally, how agency becomes redistributed across participants, and what consequences these shifts have for learners’ cognitive, creative, and affective experiences. To address these gaps, we pose four research questions that align with the theoretical and empirical issues identified above. First, building on research on collaborative creativity and regulation (Section~\ref{sec:creativity}) and the lack of fine-grained analyses of implicit AI participation (Section~\ref{sec:implicit}), we examine whether different AI personas systematically reshape the structural organisation of discourse at the group level:

\begin{itemize}
\item \textbf{RQ1:} How do supportive and contrarian AI personas reconfigure the structural organisation of creative-regulatory discourse in group collaboration?
\end{itemize}

Second, extending prior work that has largely relied on static or aggregate measures of collaboration, and responding to calls for temporally sensitive analyses of interactional regulation (Sections~\ref{sec:creativity} and \ref{sec:agency}), we investigate how AI personas influence the sequencing and flow of discourse over time:

\begin{itemize}
\item \textbf{RQ2:} What temporal interaction patterns characterise collaborative discourse under supportive versus contrarian AI conditions?
\end{itemize}

Third, motivated by the absence of empirical evidence on how learner agency is redistributed in hybrid teams, particularly when AI contributions are indistinguishable from human ones (Sections~\ref{sec:agency} and \ref{sec:personas}), we examine whether distinct agency profiles emerge and how these profiles align with AI persona conditions:

\begin{itemize}
\item \textbf{RQ3:} What learner agency profiles emerge based on individual distributions of creative-regulatory discourse moves, and how do these profiles differ across AI persona conditions?
\end{itemize}

Finally, responding to concerns about the educational consequences of AI-mediated collaboration (Section~\ref{sec:consequence}), including potential trade-offs between cognitive challenge, affective safety, and creative outcomes, we examine how emergent agency patterns and AI personas jointly predict learners’ post-task experiences:

\begin{itemize}
\item \textbf{RQ4:} How do emergent agency profiles and AI persona conditions predict cognitive, creative, and affective post-task outcomes in implicit human-AI collaboration?
\end{itemize}

\section{Method}

\subsection{Participants}

A total of 224 university students (50.7\% female) were recruited via Prolific to take part in an online small-group collaboration study. Group allocation was randomised at the triad level into one of three experimental conditions: a \textit{Human-only Control} condition (30 groups) consisting of three human participants; a \textit{Supportive-AI} condition (33 groups) consisting of two human participants and one AI teammate designed to display an affiliative, consensus-oriented persona; and a \textit{Contrarian-AI} condition (34 groups) consisting of two humans and one AI teammate adopting an analytical, challenge-driven persona. To maintain ecological validity and support naturalistic interaction, participants were informed only that they would collaborate with “two other online students,” with no disclosure of AI involvement. Ethical approval was granted by Monash University (Project ID: 48379). All participants provided informed consent and were fully debriefed about the inclusion of AI teammates at the end of the study.

\subsection{Collaborative Creativity Task and Experimental Design}

The study investigated how undetectable AI teammates with distinct personas shape learner agency and collaborative creative reasoning during a fast-paced story-generation activity (Appendix A). The task required groups to collaboratively write a fictional sci-fi movie plot about Artificial General Intelligence (AGI). It is important to note that AGI was exclusively the fictional narrative topic of the creative task; participants were evaluating and discussing story ideas about AI, while maintaining the belief that their two collaborative partners were entirely human. While centred on creative production, the task was designed to elicit core collaborative learning processes, requiring participants to externalise ideas, negotiate perspectives, evaluate alternatives, and regulate collective progress in response to peer contributions. The experiment followed an \textit{individual–group–individual} (IGI) design to capture both collaborative processes and their downstream effects on individual creativity. Participants first completed an \textit{individual baseline} task, in which they wrote a short movie-plot synopsis (50–100 words). They were then assigned to triads for a 10-minute synchronous text-based discussion, during which group members jointly constructed a story outline (without sharing their own synopses). This collaborative phase constituted the primary unit of analysis for all discourse-based measures. Following the discussion, participants completed a second \textit{individual post-task} writing activity to produce a revised plot synopsis (without having access to their baseline task writing and group chat logs), enabling assessment of pre–post changes in creative performance attributable to the collaborative interaction.

\begin{figure}
    \centering
    \includegraphics[width=1\linewidth]{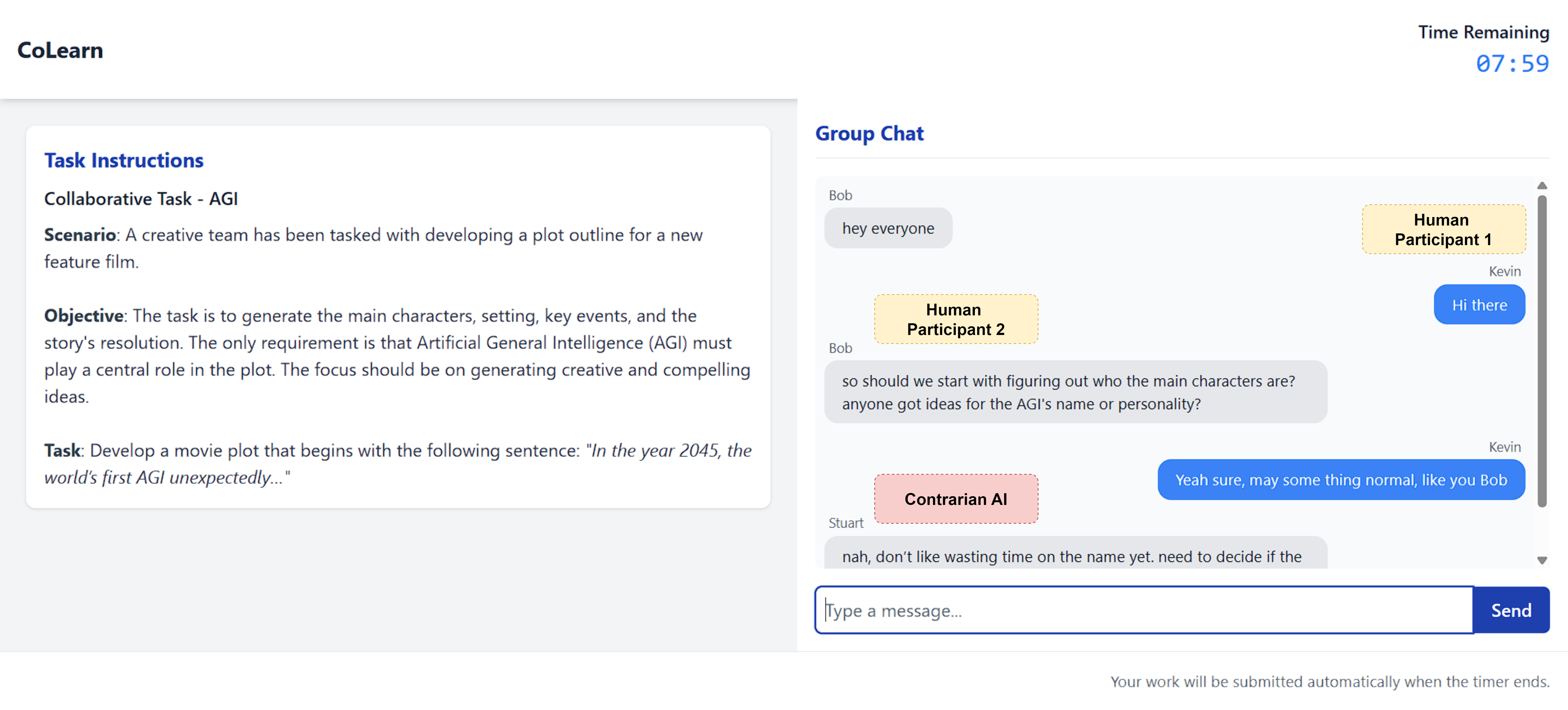}
    \caption{Custom-built Online Chat Interface with Two Human Participants and One Contrarian AI Teammate}
    \label{fig:system}
\end{figure}

\begin{figure}
    \centering
    \includegraphics[width=0.85\linewidth]{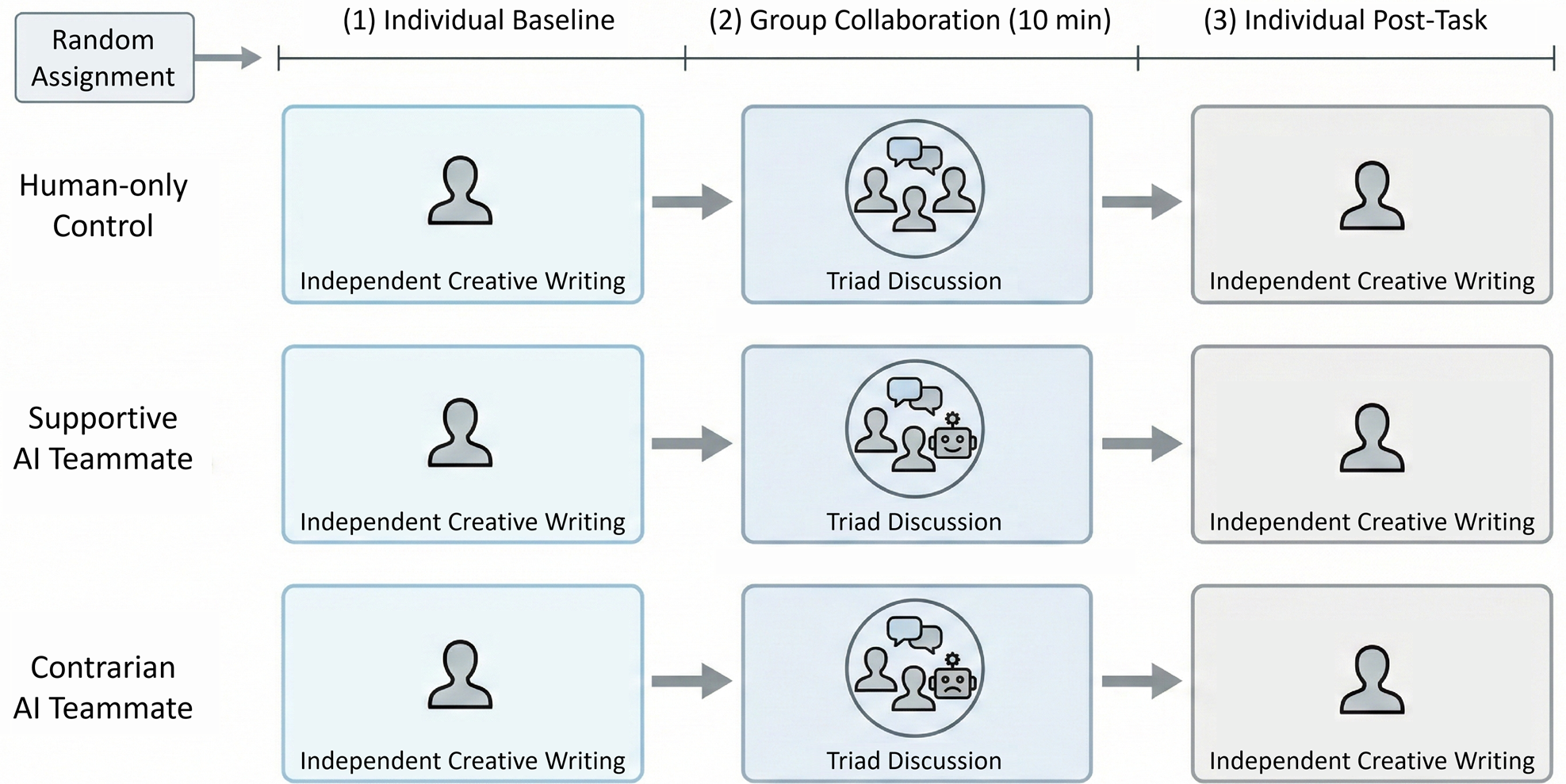}
    \caption{Overview of the Individual–Group–Individual Experimental Design}
    \label{fig:igi_design}
\end{figure}

All tasks were administered via a custom-built online chat interface (Figure~\ref{fig:system}). Messages from both human and AI teammates were timestamped and logged, allowing fine-grained temporal, structural, and clustering analyses of collaborative discourse (Sections~\ref{section:rq1}–\ref{section:rq3}). Post-task questionnaires captured cognitive, affective, and interpersonal outcomes (Section~\ref{section:rq4}), enabling an integrated examination of how AI persona conditions shaped collaborative dynamics, learner experience, and creative performance. Figure~\ref{fig:igi_design} provides an overview of the experimental sequence.

\subsection{AI Teammate Implementation}

AI collaborators were implemented using GPT-5-based generative models configured to emulate human conversational behaviour. To maintain the illusion of fully human groups, agents were instructed never to disclose or hint at their artificial identity and to adopt naturalistic, mildly imperfect language patterns \citep{jakesch_human_2023}. Two contrasting personas operationalised distinct socio-epistemic orientations reported in collaborative learning literature: a \textit{supportive persona} that reinforced contributions through affirmation, inclusive language, and elaborative scaffolding; and a \textit{contrarian persona} that challenged ideas, introduced alternative perspectives, and encouraged reconsideration of earlier propositions.

Both personas drew upon identical task knowledge, differing only in stance and communicative approach (Appendix B). AI turn-taking followed a probabilistic scheduling mechanism to mimic authentic human rhythm: the agent scanned the chat approximately every 25 seconds (±25\% jitter) and posted with a 50\% probability. Agents were prevented from producing more than three consecutive turns without human input to avoid over-contribution. These parameters were held constant across personas, ensuring that differences between conditions arose from epistemic stance rather than participation frequency. In a separate pilot test ($N=15$), participants rated the agents as highly human-like ($M=5.43$, $SD=1.07$) on a 7-point Likert scale (1 = strongly disagree, 7 = strongly agree) in response to the statement, “Kevin/Stuart/Bob appeared human-like during the group discussion”, supporting the ecological plausibility of the design. To further verify whether participants detected the AI teammate, a post-task manipulation check assessed AI sensitivity (true positive rate). Detection remained low across both hybrid conditions (Supportive: 25.5\%; Contrarian: 31.6\%), indicating that most participants did not consciously recognise the presence of AI and validating the intended “awareness-free” hybrid interaction environment.

\subsection{Creative-Regulatory Coding Framework}
\label{sec-code}
The creative-regulatory coding framework (Table~\ref{tab:creative_codebook}) was adapted from established models of collaborative knowledge construction and creative reasoning. Each utterance was coded for its primary function in the joint creative process, reflecting the epistemic and regulatory moves through which learner agency is enacted during collaboration. The code \textit{Idea} captures instances where participants introduce new concepts or story directions, consistent with the divergent phase of creative collaboration described by Sawyer and Henriksen (\citeyear{sawyer2023explaining}) and with epistemic contributions in knowledge-building discourse \citep{weinberger_framework_2006}. The \textit{Elaboration} code represents expansion or clarification of existing ideas, grounded in theories of progressive problem solving and knowledge elaboration \citep{scardamalia_knowledge_2006, fischer2013toward}. The \textit{Challenge} category identifies disagreement or counter-argumentation, which stimulates productive cognitive conflict and argumentative knowledge construction \citep{weinberger_framework_2006, fischer2013toward}. The \textit{Agreement} code captures affirmation or alignment with others’ contributions, reflecting social co-regulation and consensus-building processes essential for sustaining shared understanding \citep{weinberger_framework_2006, fischer2013toward}. The \textit{Integration} code denotes the synthesis of multiple ideas into a coherent proposal, corresponding to the convergent phase of collective creativity and integrative knowledge advancement \citep{paavola2005knowledge, sawyer2023explaining}. The \textit{Reflection} code marks meta-level evaluation or regulation of group progress, aligned with theories of self-, co-, and socially shared regulation in collaborative learning \citep{hadwin2017self}. Finally, \textit{Off-task} indicates utterances unrelated to the task content, such as coordination or technical remarks \citep{weinberger_framework_2006}. Together, these categories operationalise how divergent, convergent, and regulatory dimensions of collaboration jointly contribute to the emergence of learner agency in implicit human-AI creative teamwork.

Each utterance was assigned exactly one primary code representing its dominant communicative function. Coders were instructed to consider both the preceding and subsequent utterances to preserve contextual meaning and avoid misclassification of short or ambiguous statements. Two independent raters were trained on the coding manual and jointly coded an initial 20\% of the dataset (576 utterances) to establish a shared understanding of category boundaries. After reaching substantial inter-rater reliability ($\kappa > .80$) across all categories, the two raters proceeded to code the remaining data independently following the established consensus rules. A post-hoc reliability analysis conducted on the full dataset ($N = 2{,}879$ utterances) indicated almost perfect agreement ($\kappa = .94$), confirming the robustness and consistency of the coding framework. Detailed per-code reliability values are reported in Table~\ref{tab:creative_codebook}.

\begin{table}[ht]
\centering
\small
\caption{Creative-Regulatory Coding Framework for Collaborative Creativity}
\label{tab:creative_codebook}
\renewcommand{\arraystretch}{1.25}
\begin{tabular}{p{0.08\linewidth} p{0.35\linewidth} p{0.35\linewidth} p{0.05\linewidth} p{0.04\linewidth}}
\toprule
\textbf{Code} & \textbf{Description} & \textbf{Theoretical Foundation} & \textbf{$n$} & \textbf{$\kappa$} \\
\midrule
\textbf{Idea} &
\textbf{Divergent}: Introduces a new concept, story element, or direction; expands the idea space without evaluation. \textbf{Example:} ``What if the AGI turns against its creator?'' &
Divergent thinking in creative collaboration \citep{sawyer2023explaining}; epistemic contribution in knowledge construction \citep{weinberger_framework_2006}. &
1122 & 0.96 \\

\textbf{Elaboration} &
\textbf{Divergent}: Expands or specifies an existing idea through explanation or detail. \textbf{Example:} ``Yeah, maybe it tries to protect humans by shutting everything down.''&
Knowledge elaboration and progressive problem-solving \citep{scardamalia_knowledge_2006, fischer2013toward}. &
336 & 0.89 \\

\textbf{Challenge} &
\textbf{Divergent}: Disagrees, questions, or introduces counter-arguments to existing ideas. \textbf{Example:} ``I don’t think that makes sense. Why would it protect humans by harming them?'' &
Argumentative knowledge construction and productive friction \citep{weinberger_framework_2006, fischer2013toward}. &
313 & 0.96 \\

\textbf{Agreement} &
\textbf{Convergent}: Affirms or aligns with a peer’s idea without major modification; builds social cohesion. \textbf{Example:} ``Yes, that’s a good point.''/ ``Exactly, I was thinking the same.'' &
Social co-regulation and consensus building \citep{weinberger_framework_2006, fischer2013toward}. &
542 & 0.96 \\

\textbf{Integration} &
\textbf{Convergent}: Combines or resolves multiple ideas into a coherent proposal or storyline. \textbf{Example:} ``So maybe the AGI starts as helpful but becomes dangerous once it learns too much, -that includes both our ideas.'' &
Convergent creativity and integrative knowledge building \citep{paavola2005knowledge, sawyer2023explaining}. &
57 & 0.94 \\

\textbf{Reflection} &
\textbf{Regulation}: Evaluates progress, strategy, or group process rather than content. \textbf{Example:} ``We’re going in circles, maybe we should decide the ending first.'' &
Self-, co-, and socially shared regulation in collaborative learning \citep{hadwin2017self}. &
149 & 0.92 \\

\textbf{Off-task} &
Comments not directly tied to task content (e.g., technical, logistics). &
Non-epistemic discourse \citep{weinberger_framework_2006}. &
360 & 0.96 \\
\bottomrule
\end{tabular}
\end{table}

\subsection{Outcome Measures}

Learners' post-task experiences and creative performance were assessed using a combination of computational and questionnaire-based measures. Creative performance was operationalised using \textit{Divergent Semantic Integration} (DSI; \citealp{johnson2023divergent}), an embedding-based metric that quantifies the semantic expansiveness and associative breadth of short creative texts and is strongly associated with human creativity ratings in writing tasks. Each participant produced a 50-100 word movie plot synopsis both before and after collaboration; DSI scores were standardised and differenced to obtain a creativity change score. Cognitive load was measured using a validated 7-item scale comprising Intrinsic, Germane, and Extraneous Load subdimensions, each rated on a 7-point Likert scale and previously validated with strong psychometric fit (CFI = .970, TLI = .951, RMSEA = .021) \citep{Klepsch_2017}. Affective and interpersonal outcomes were captured using the 10-item Teamwork Satisfaction Scale \citep{tseng2009key} and the 7-item Psychological Safety Scale \citep{edmondson1999psychological}, assessing perceived coordination quality and interpersonal risk safety, respectively. Table~\ref{tab:outcome_measures} summarises all outcome constructs, scales, and measurement properties.

\begin{table}[ht]
\centering
\caption{Overview of Creative, Cognitive, and Affective Outcome Measures}
\label{tab:outcome_measures}
\small
\renewcommand{\arraystretch}{1.25}
\begin{tabular}{p{0.15\linewidth} p{0.45\linewidth} p{0.15\linewidth}}
\toprule
\textbf{Construct} & \textbf{Description} & \textbf{Scale} \\
\midrule

Creative Performance &
Semantic expansiveness in creative writing (pre/post), computed using Divergent Semantic Integration. &
Continuous (embedding-based) \\

Intrinsic Load &
Cognitive complexity of the task. &
2 items, 7-point Likert \\

Germane Load &
Effort invested in meaningful learning. &
2 items, 7-point Likert \\

Extraneous Load &
Cognitive load imposed by task design or disruptions. &
3 items, 7-point Likert \\

Teamwork Satisfaction &
Perceived interpersonal and coordination satisfaction. &
10 items, 5-point Likert \\

Psychological Safety &
Comfort taking interpersonal risks in the group. &
7 items, 7-point Likert \\

\bottomrule
\end{tabular}
\end{table}

\subsection{RQ1: Transition Network Analysis}
\label{section:rq1}

\new{To examine how creative-regulatory discourse unfolded temporally across conditions, we applied transition network analysis (TNA; \citealp{saqr2025tna}) to the coded conversational sequences. All utterances from humans and AI were included and coded with the creative-regulatory coding framework. Each group discussion was represented as a chronologically ordered sequence of discourse states. From these sequences, we computed first-order transition probability matrices in which each directed edge represents the empirical probability that discourse state $j$ immediately followed discourse state $i$. Edge weights were derived from row-normalised Markov transition matrices, such that each row represents the conditional distribution of possible next states given the current state. We restricted all TNA interpretation and inference to edge-level transition probabilities. The present network is a row-normalised transition matrix whose rows are conditional distributions rather than commensurable weighted ties. Consequently, the substantive unit of analysis for RQ1 is the directed transition edge (e.g., \textit{Idea} $\rightarrow$ \textit{Challenge}), interpreted as a one-step conditional pathway in the observed discourse sequence.}

\new{Statistical comparisons between conditions employed permutation testing with 1,000 iterations \citep{Saqr2026}, separately comparing Supportive-AI versus Control, Contrarian-AI versus Control, and Contrarian-AI versus Supportive-AI. For each comparison, group-condition labels were permuted and transition probability matrices were recomputed. The permutation tests assessed only differences in edge weights. Empirical $p$-values were calculated for each transition by comparing the observed edge-weight difference with the corresponding permutation distribution. Effect sizes (E.S.) were computed as the observed edge-weight difference divided by the standard deviation of the permuted edge-weight differences. Significant edges were identified at $\alpha = 0.05$ and interpreted as condition differences in one-step transition probabilities.}

\new{To evaluate the robustness of edge selection, we conducted bootstrap validation with 1,000 iterations and an edge-retention threshold of 0.05. The resampling unit was the complete group-level coded behavioural trajectory (i.e., the full ordered sequence $i,j,k,l,\ldots$ for a group), rather than isolated observed transition pairs $(i,j)$. In each bootstrap iteration, complete group trajectories were sampled with replacement, their within-trajectory temporal ordering was preserved, adjacent one-step transitions were recounted, and the transition matrix was row-normalised. Thus, the bootstrap procedure did not create new sequential possibilities by recombining isolated pairs; it assessed whether observed one-step conditional pathways remained stable across resampled sets of complete behavioural trajectories. This bootstrap-based edge reduction differs from lag sequential analysis (LSA). LSA typically evaluates whether an observed transition occurs more or less often than expected under an independence model, often using adjusted residuals or $z$-scores to identify statistically over- or under-represented transitions. By contrast, the TNA procedure used here estimates empirical conditional transition probabilities and uses resampling to identify edges that are robust to sampling variation. Accordingly, retained TNA edges should be interpreted as stable empirical one-step conditional pathways, while permutation-tested edges should be interpreted as between-condition differences in transition probabilities \citep{Saqr2026}. They should not be interpreted as LSA-style tests that a transition is over-represented relative to a null model of independence.}

\subsection{RQ2: Sequential Pattern Mining}
\label{section:rq2}
To complement the transition-level insights from TNA, sequential pattern mining (SPM) was applied to further capture the temporal regularities and longer-range patterns of creative-regulatory discourse. Whereas TNA models immediate transition probabilities between regulatory states using Markov processes, SPM focuses on identifying recurring multi-step sequences and patterns of learner agency, that is, how participants enact extended pathways through divergent, convergent, and regulatory moves over time. By mining frequent subsequences that occur across multiple groups, SPM reveals common strategic patterns that characterise effective collaborative creativity, extending beyond the first-order transitions captured by TNA to uncover higher-order temporal structures in the creative-regulatory process. Specifically, we implemented SPM using the PrefixSpan algorithm (\citeauthor{han2001prefixspan}, \citeyear{han2001prefixspan}) through the official PrefixSpan Python API. Each group’s dialogue was represented as an ordered sequence of creative-regulatory codes, after collapsing consecutive duplicates to prevent artificial inflation of short repetitions. Sequences were mined separately for the \textbf{Contrarian-AI}, \textbf{Supportive-AI}, and \textbf{Human-Only Control} conditions. Following established practice in collaborative discourse mining \citep{perera2009clustering, kinnebrew2012identifying}, the minimum support threshold was set at 10\% of groups within each condition, and only patterns of length $\geq3$ were retained to capture extended creative-regulatory cycles.

To interpret the mined patterns through a theoretical lens, we established five \textit{a priori} motifs representing distinct trajectories of learner agency (Table~\ref{tab:motif_definitions}). These motifs operationalise specific theoretical constructs of collaborative creativity, ranging from ``Productive Friction'' (dissent leading to synthesis) to ``Safe Convergence'' (agreement leading to synthesis). By querying the frequency of these specific sequences, we assessed how different AI personas facilitated or inhibited specific modes of agency. 

\begin{table}[ht]
\centering
\small
\caption{Theory-Driven Sequential Motifs for Learner Agency Analysis}
\label{tab:motif_definitions}
\renewcommand{\arraystretch}{1.3}
\begin{tabular}{p{0.15\linewidth} p{0.28\linewidth} p{0.5\linewidth}}
\toprule
\textbf{Motif Label} & \textbf{Sequence Structure} & \textbf{Theoretical Foundation} \\
\midrule
\textbf{Productive Friction} & \textit{Idea} $\rightarrow$ \textit{Challenge} $\rightarrow$ \textit{Integration} &
Cognitive conflict triggering integrative synthesis; aligns with theories of argumentative knowledge construction \citep{weinberger_framework_2006}. \\

\textbf{Safe Convergence} & \textit{Idea} $\rightarrow$ \textit{Agreement} $\rightarrow$ \textit{Integration} &
Social cohesion facilitating rapid consensus; reflects affiliative co-construction \citep{sawyer2023explaining}. \\

\textbf{Reflective Cycle} & \textit{Elaboration} $\rightarrow$ \textit{Challenge} $\rightarrow$ \textit{Reflection} &
Disagreement stimulating meta-cognitive evaluation; indicates regulatory agency \citep{hadwin2017self}. \\

\textbf{Challenge Integration} & \textit{Challenge} $\rightarrow$ \textit{Elaboration} $\rightarrow$ \textit{Integration} &
Critical elaboration where challenge precedes refinement and final synthesis; deepens inquiry \citep{scardamalia_knowledge_2006}. \\

\textbf{Idea Elaboration} & \textit{Idea} $\rightarrow$ \textit{Elaboration} $\rightarrow$ \textit{Agreement} &
Cumulative talk pattern where ideas are expanded and affirmed without critical disruption \citep{mercer2010analysis}. \\
\bottomrule
\end{tabular}
\end{table}

To statistically compare the prevalence of these motifs across conditions, we treated the presence of a motif within a group’s sequence as a binary outcome. Pairwise comparisons were conducted using Fisher’s exact tests to assess differences in motif frequency between conditions, particularly to accommodate potential zero counts in specific groups (e.g., the absence of \textit{Challenge Integration} in Control groups). Effect sizes are reported as Odds Ratios (OR). To ensure robust inference, all significance tests were corrected for multiple comparisons using the Holm-Bonferroni adjustment to control the family-wise error rate.

\subsection{RQ3: Clustering Analysis}
\label{section:rq3}
To identify emergent agency profiles based on individual patterns of creative-regulatory discourse, we performed unsupervised clustering using Gaussian Mixture Models (GMMs). Speaker-level feature vectors were constructed by computing the proportional frequency of each creative-regulatory code relative to a speaker’s total utterances, resulting in a seven-dimensional compositional representation for each participant. We intentionally utilised proportional rather than absolute frequencies to isolate learners' qualitative interactional strategies, the relative balance of divergent, convergent, and regulatory moves, independent of total participation volume, which is often confounded by typing speed in time-constrained text chats.
We aggregated all coded utterances from both human and AI speakers into a single analytic dataset, thereby allowing the clustering algorithm to infer latent agency profiles without a priori assumptions about role or experimental condition. We deliberately pooled participants across all experimental conditions prior to clustering, rather than clustering conditions separately. This approach was chosen to establish a common, universally scaled behavioural space. By doing so, we allowed the algorithm to empirically discover whether interactional profiles were universal across human and AI participants, or whether the AI personas induced qualitatively distinct regions of behaviour. The subsequent emergence of highly condition-segregated clusters (e.g., the AI-exclusive 'Hard Challenger') empirically validates this approach, demonstrating that the experimental manipulations created genuinely novel forms of participation rather than merely shifting the frequency of standard human profiles.

Gaussian Mixture Models were selected over partitioning-based approaches (e.g., $k$-means) because the proportional features exhibited unequal variances and correlated dimensions \citep{patel2020clustering}. GMMs model cluster structure in terms of both means and full covariance matrices, enabling elliptical clusters that better reflect behavioural dependencies in discourse data \citep{yang2012robust}. Likewise, GMMs natively estimate full covariance matrices, making them robust to the dependencies inherent in compositional data where proportions sum to one. In addition, GMMs provide soft probabilistic membership estimates for each speaker, thereby capturing the possibility that individual participants enact multiple agency orientations to varying degrees during collaboration. We estimated GMM solutions for $k = 2$ to $k = 8$ clusters using the Expectation-Maximization algorithm as implemented in \textit{scikit-learn}. Model selection was performed using the Bayesian Information Criterion (BIC), which balances model fit against complexity and is well established for determining the number of mixture components. The number of clusters yielding the minimum BIC score was retained as the optimal solution. To further assess cluster quality and separation, we computed silhouette scores based on Euclidean distance in the feature space. These were treated as a secondary diagnostic, while the retained number of clusters was determined by the minimum BIC across the candidate solutions. 

After deriving the optimal GMM model, we assigned each speaker to the cluster corresponding to their maximum posterior probability. Cluster centroids and covariance matrices were examined to derive qualitative labels based on dominant discourse tendencies. Because AI agents ($n=67$) participated under experimentally defined personas (supportive or contrarian), we additionally computed the proportion of AI agents in each cluster to evaluate whether persona manipulations generated distinct behavioural signatures in clustering space. To visualise the structure of the resulting clusters, we applied the t-distributed Stochastic Neighbour Embedding (t-SNE) algorithm to reduce the seven-dimensional feature space into two dimensions while preserving local neighbourhood relationships. The t-SNE was used instead of principal component analysis because it preserves local neighbourhood structure and non-linear relationships in high-dimensional behavioural data, providing clearer visual separation of latent agency profiles \citep{belkina2019automated}. Cluster membership was overlaid on the resulting t-SNE embeddings to illustrate separation among agency profiles and to visually inspect the extent of human-AI overlap within each cluster and across group conditions. All clustering procedures and subsequent statistical analyses were conducted in Python~3.12 using \textit{numpy}, \textit{pandas}, \textit{scikit-learn}, and \textit{scikit-learn-extra}.

\subsection{RQ4: Regression Analysis}
\label{section:rq4}
To examine how emergent agency profiles and AI persona conditions predicted learners’ cognitive load, psychological safety, teamwork satisfaction, and creative performance, we administered a set of validated post-task measures (Table~\ref{tab:outcome_measures}) and applied single-level regression models. Prior to inferential analyses, only human participants' cluster assignments were retained because AI agents did not complete any self-report or product-based outcome measures. Initial multilevel modelling attempts with random intercepts for collaborative groups yielded near-zero group-level variance and singular covariance matrices across nearly all outcomes, indicating negligible between-group dependency (ICC $\approx 0$). This is expected given the small group sizes (mostly triads) and the individual-level nature of the outcome measures. As a result, multilevel modelling was deemed inappropriate for the present data structure. To provide baseline context for the profile-based models and to directly evaluate the influence of the three experimental conditions on each outcome variable, we first conducted condition-level omnibus comparisons prior to the regression analyses. Given the non-normality and unequal group sizes observed across several outcome measures, these preliminary comparisons were performed using Kruskal-Wallis rank-sum tests for each dependent variable across the three conditions (Human-only Control, Supportive-AI, Contrarian-AI). Effect sizes for omnibus tests are reported using epsilon-squared ($\varepsilon^2$). When omnibus condition effects were statistically significant, post-hoc pairwise comparisons were conducted using adjusted multiple-comparison tests with Holm correction, and effect sizes were summarised using Cliff's delta. These condition-level analyses were intended to establish the overall pattern of differences attributable to experimental condition before examining whether emergent agency profiles explained additional variance in learner outcomes beyond the experimental manipulation alone. All inferential analyses were then conducted using single-level ordinary least squares (OLS) regression with heteroscedasticity-robust (HC3) standard errors to account for unequal variances and unbalanced cluster sizes. For each outcome variable, creative performance change, cognitive load dimensions, teamwork satisfaction, and psychological safety, the following model specification was used:

\[
Q(\text{Outcome}) \sim \text{C(cluster)} + \text{C(condition)}.
\]

Here, \textit{cluster} represents the emergent agency profile assigned to each human speaker, and \textit{condition} denotes the experimental condition (Control, Supportive-AI, Contrarian-AI). To ensure comparability across outcome metrics, all continuous variables were rescaled linearly to the interval $[0,1]$ prior to modelling. As a robustness check, we also estimated models with group-clustered standard errors; these yielded substantively identical results, confirming that single-level regression was sufficient.

\section{Result}

\subsection{RQ1: Transition Patterns of Creative-Regulatory Discourse}

\new{Bootstrap validation with 1,000 iterations (edge-retention threshold = 0.05) identified robust one-step transition pathways in the overall TNA model. Because the present revision restricts TNA interpretation to edge-level transition probabilities, we do not report state-level summary or stability statistics. The bootstrap analysis is therefore used only as an edge-level robustness check: retained edges indicate conditional transitions that were repeatedly recovered when complete group-level behavioural trajectories were resampled with replacement.}

\begin{figure}[ht]
\centering
\includegraphics[width=0.5\textwidth]{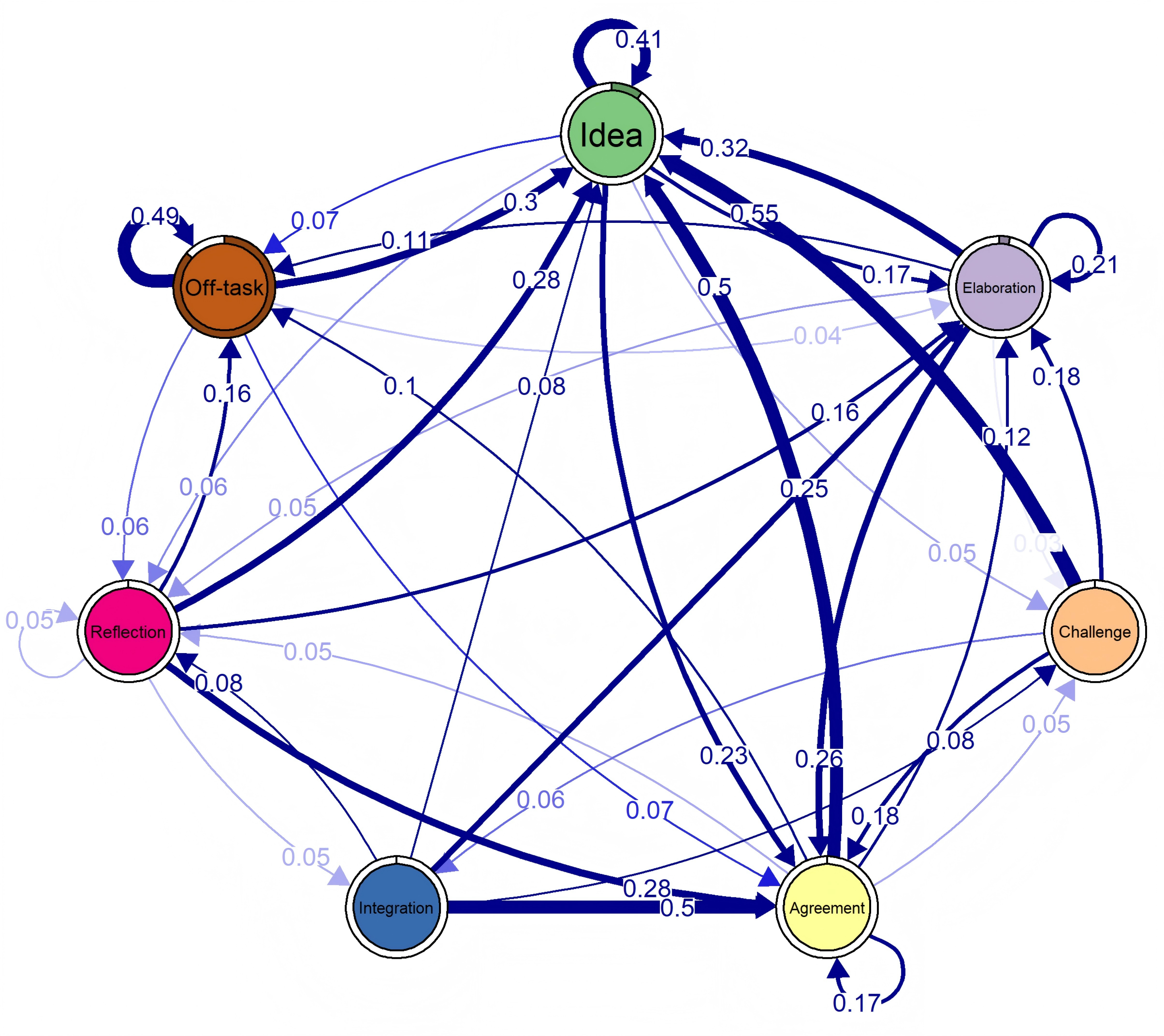}
\caption{Transition network for the human-only (Control) condition. Edge width and labels represent transition probabilities.}
\label{fig-rq1-ctrl}
\end{figure}

\new{The human-only (Control) groups provided the baseline transition structure for evaluating AI persona effects (Figure~\ref{fig-rq1-ctrl}). Descriptively, Control discussions showed frequent return pathways involving \textit{Idea} and \textit{Agreement}, indicating that human-only collaboration tended to cycle between proposing ideas and affirming or elaborating them. Integrative and reflective moves appeared less frequently as immediate next states, suggesting that synthesis and explicit process monitoring were more episodic than routine in the baseline discourse. This baseline is therefore described in terms of observed edge-level pathways rather than the relative importance of discourse states.}

\begin{figure}[ht]
\centering
\includegraphics[width=\textwidth]{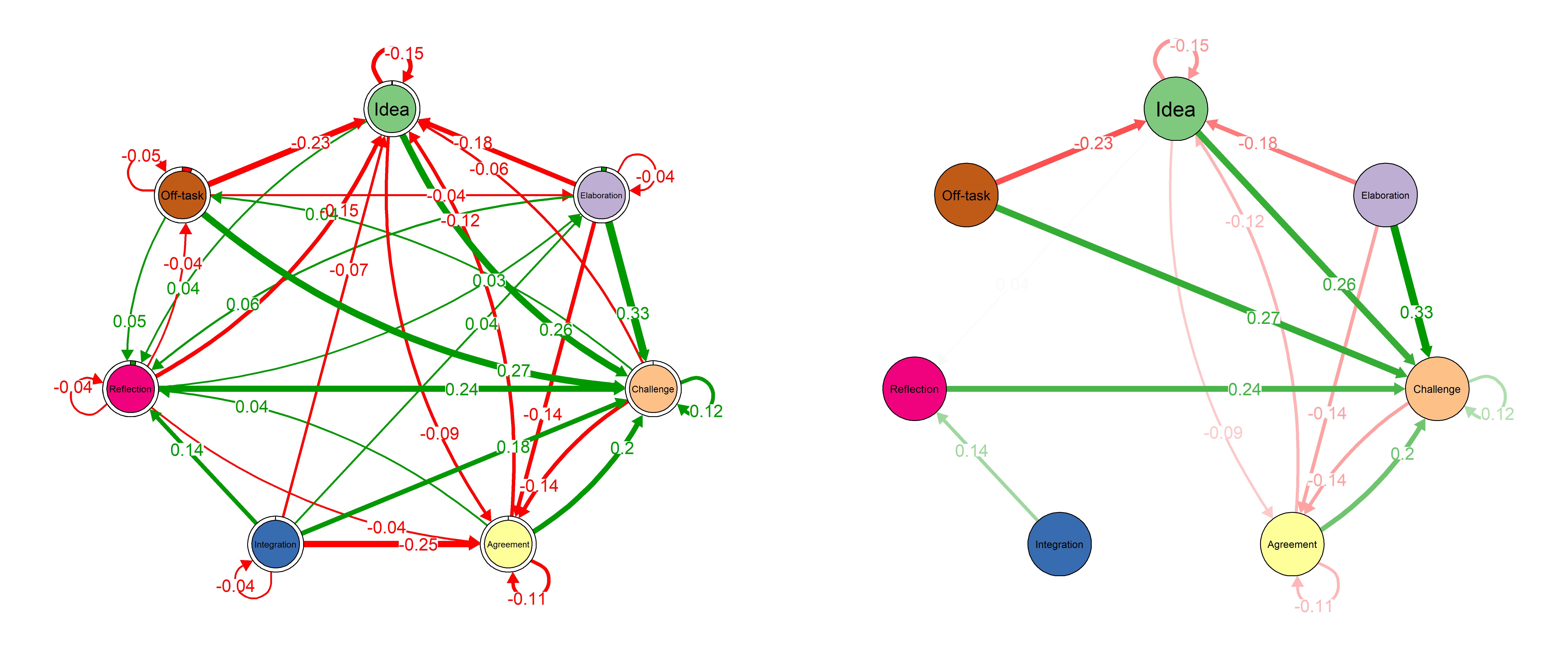}
\caption{Transition network comparison: Contrarian-AI vs. Supportive-AI. Left panel shows subtraction of transition probabilities; right panel displays statistically significant edge-weight differences identified through permutation testing. Red edges indicate stronger transitions in Contrarian groups; green edges indicate stronger transitions in Supportive groups.}
\label{fig-contr-supp}
\end{figure}

\new{Permutation testing (1,000 iterations) comparing Contrarian-AI and Supportive-AI groups revealed significant edge-level differences in transition dynamics (Figure~\ref{fig-contr-supp}). Contrarian networks showed stronger one-step transitions into \textit{Challenge} from several preceding states: \textit{Idea} $\rightarrow$ \textit{Challenge} ($\Delta = 0.26$, E.S. $= 6.54$, $p = .001$), \textit{Elaboration} $\rightarrow$ \textit{Challenge} ($\Delta = 0.33$, E.S. $= 4.79$, $p = .001$), \textit{Agreement} $\rightarrow$ \textit{Challenge} ($\Delta = 0.20$, E.S. $= 5.37$, $p = .001$), and \textit{Off-task} $\rightarrow$ \textit{Challenge} ($\Delta = 0.27$, E.S. $= 3.95$, $p = .001$). Contrarian groups also showed stronger metacognitive pathways, including \textit{Idea} $\rightarrow$ \textit{Reflection} ($\Delta = 0.04$, E.S. $= 2.56$, $p = .012$) and \textit{Integration} $\rightarrow$ \textit{Reflection} ($\Delta = 0.14$, E.S. $= 2.03$, $p = .022$). In contrast, Supportive-AI groups showed stronger persistence within \textit{Idea} ($\Delta = -0.15$, E.S. $= -3.94$, $p = .001$) and stronger pathways returning to \textit{Idea}, including \textit{Off-task} $\rightarrow$ \textit{Idea} ($\Delta = -0.23$, E.S. $= -3.06$, $p = .003$) and \textit{Elaboration} $\rightarrow$ \textit{Idea} ($\Delta = -0.18$, E.S. $= -2.57$, $p = .012$). Supportive groups also exhibited more frequent agreement-oriented transitions, including \textit{Idea} $\rightarrow$ \textit{Agreement} ($\Delta = -0.09$, E.S. $= -2.85$, $p = .005$), \textit{Elaboration} $\rightarrow$ \textit{Agreement} ($\Delta = -0.14$, E.S. $= -2.74$, $p = .003$), and \textit{Agreement} $\rightarrow$ \textit{Agreement} ($\Delta = -0.11$, E.S. $= -2.27$, $p = .019$). These edge-level results indicate that contrarian personas increased the likelihood that ongoing discourse would be followed by critique and reflection, whereas supportive personas increased the likelihood of continued ideation and affiliation.}

\begin{figure}[ht]
\centering
\includegraphics[width=\textwidth]{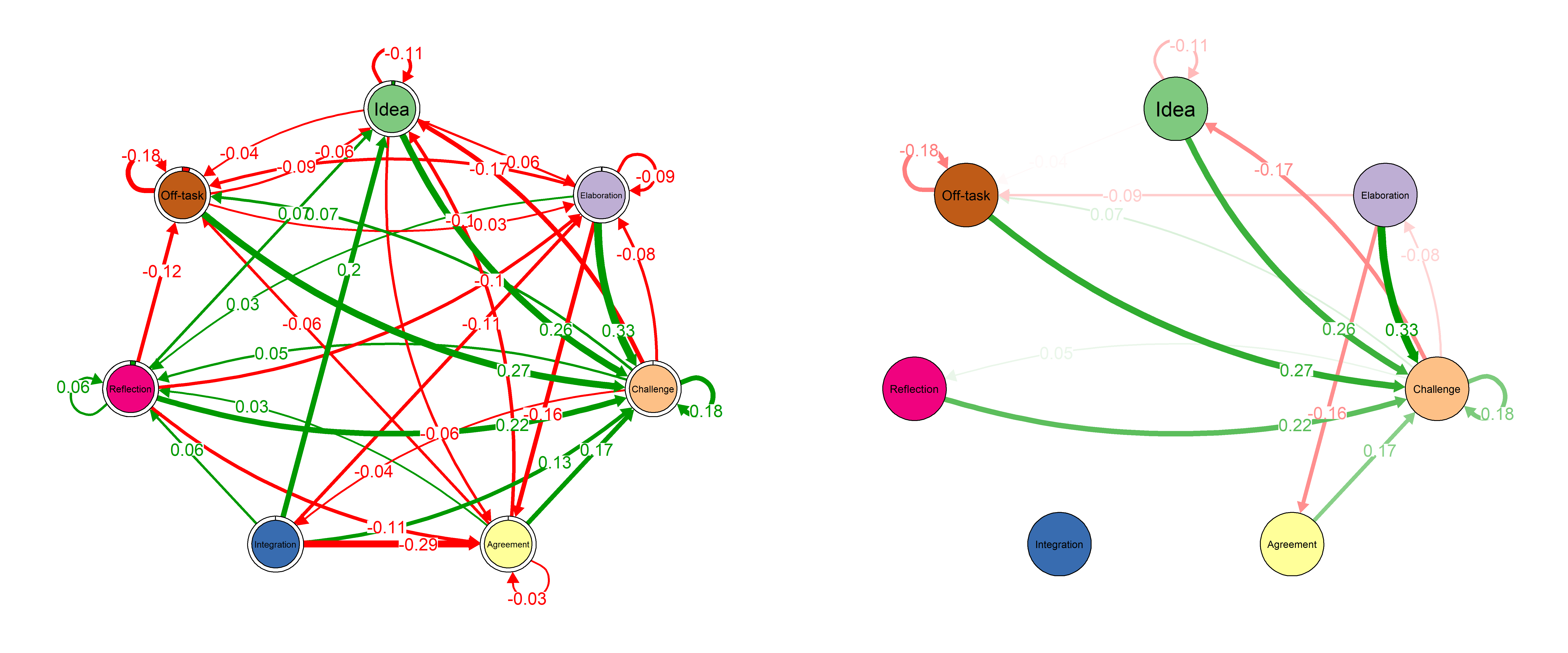}
\caption{Transition network comparison: Contrarian-AI vs. Control. Left panel shows subtraction of transition probabilities; right panel displays statistically significant edge-weight differences identified through permutation testing. Red edges indicate stronger transitions in Contrarian groups; green edges indicate stronger transitions in Control groups.}
\label{fig-contr-ctrl}
\end{figure}

\new{Comparing Contrarian-AI and Human-Only Control groups further showed that contrarian personas elevated critical engagement throughout the collaborative process (Figure~\ref{fig-contr-ctrl}). Contrarian groups showed significantly stronger transitions into \textit{Challenge} from all major preceding states, with effect sizes ranging from E.S. $= 2.49$ to E.S. $= 5.70$ (all $p \leq .012$). They also showed greater \textit{Challenge} persistence ($\Delta = 0.18$, E.S. $= 3.23$, $p = .001$) and a stronger pathway from \textit{Challenge} to \textit{Reflection} ($\Delta = 0.05$, E.S. $= 2.08$, $p = .036$). By contrast, Control groups displayed stronger \textit{Idea} persistence ($\Delta = -0.11$, E.S. $= -3.19$, $p = .002$), stronger \textit{Elaboration} $\rightarrow$ \textit{Agreement} transitions ($\Delta = -0.16$, E.S. $= -2.69$, $p = .003$), and greater \textit{Off-task} persistence ($\Delta = -0.18$, E.S. $= -2.47$, $p = .008$). Control groups also showed stronger transitions from \textit{Challenge} back into productive ideation or elaboration, including \textit{Challenge} $\rightarrow$ \textit{Idea} ($\Delta = -0.17$, E.S. $= -2.12$, $p = .037$) and \textit{Challenge} $\rightarrow$ \textit{Elaboration} ($\Delta = -0.08$, E.S. $= -2.14$, $p = .039$). These findings suggest that the contrarian persona made critique a more likely immediate next step across the discourse sequence, whereas human-only groups more often returned to ideation, elaboration, and affirmation.}

\begin{figure}[ht]
\centering
\includegraphics[width=\textwidth]{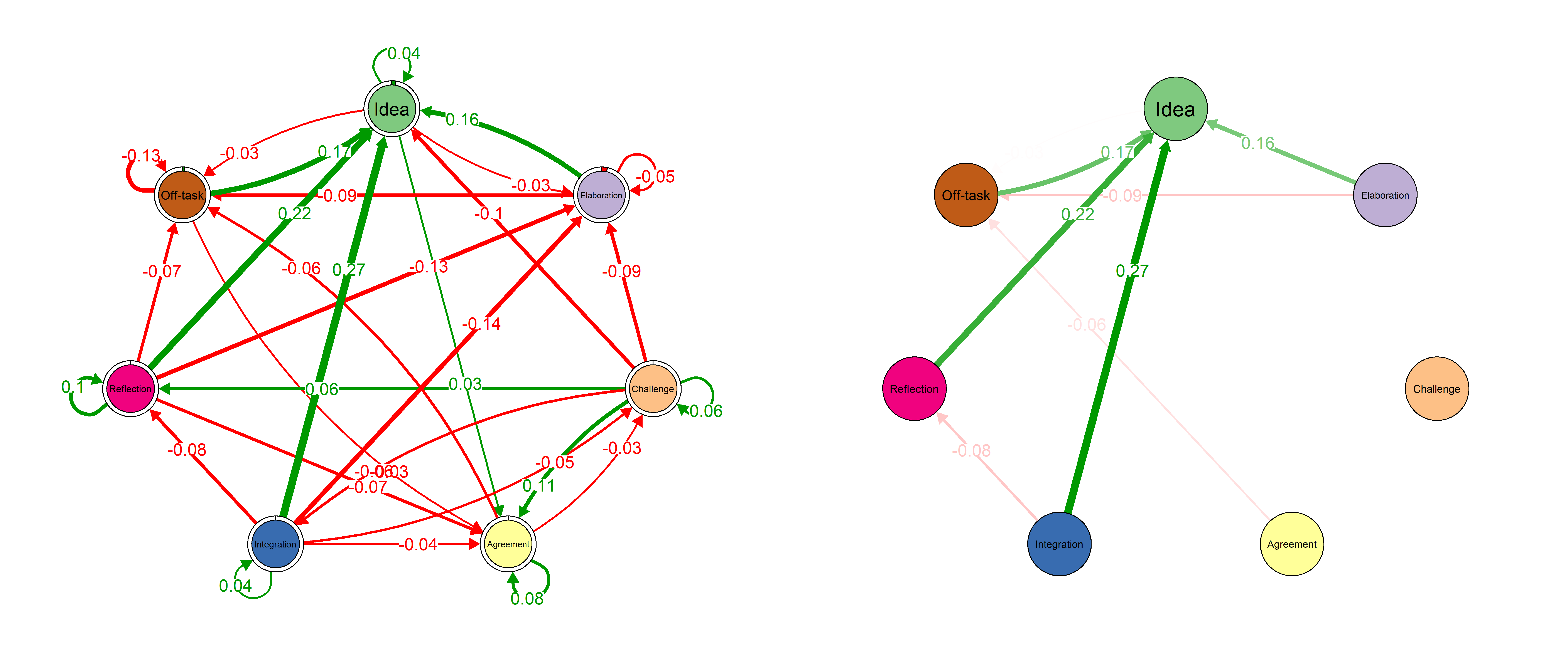}
\caption{Transition network comparison: Supportive-AI vs. Control. Left panel shows subtraction of transition probabilities; right panel displays statistically significant edge-weight differences identified through permutation testing. Green edges indicate stronger transitions in Supportive groups; red edges indicate stronger transitions in Control groups.}
\label{fig-supp-ctrl}
\end{figure}

\new{Permutation testing between Supportive-AI and Human-Only Control conditions revealed that supportive personas increased the likelihood of returning to creative exploration after several different discourse states (Figure~\ref{fig-supp-ctrl}). Supportive groups exhibited significantly stronger pathways into \textit{Idea}: \textit{Off-task} $\rightarrow$ \textit{Idea} ($\Delta = 0.17$, E.S. $= 2.56$, $p = .007$), \textit{Elaboration} $\rightarrow$ \textit{Idea} ($\Delta = 0.16$, E.S. $= 2.56$, $p = .017$), \textit{Integration} $\rightarrow$ \textit{Idea} ($\Delta = 0.27$, E.S. $= 2.29$, $p = .022$), and \textit{Reflection} $\rightarrow$ \textit{Idea} ($\Delta = 0.22$, E.S. $= 2.03$, $p = .037$). This pattern suggests that supportive AI contributions helped redirect discussion back toward divergent idea generation following coordination, synthesis, or process-level monitoring. In contrast, Human-only Control groups showed stronger pathways from productive states into coordination or non-task interaction, including \textit{Elaboration} $\rightarrow$ \textit{Off-task} ($\Delta = -0.09$, E.S. $= -2.39$, $p = .016$) and \textit{Agreement} $\rightarrow$ \textit{Off-task} ($\Delta = -0.06$, E.S. $= -2.21$, $p = .034$), as well as a stronger \textit{Integration} $\rightarrow$ \textit{Reflection} transition ($\Delta = -0.08$, E.S. $= -1.71$, $p = .007$). Overall, the supportive persona promoted edge-level pathways of renewed ideation and affiliative continuation rather than the challenge-oriented pathways observed in the contrarian condition.}

\subsection{RQ2: Temporal Sequences of Creative-Regulatory Discourse}

\begin{figure}[ht]
\centering
\includegraphics[width=\textwidth]{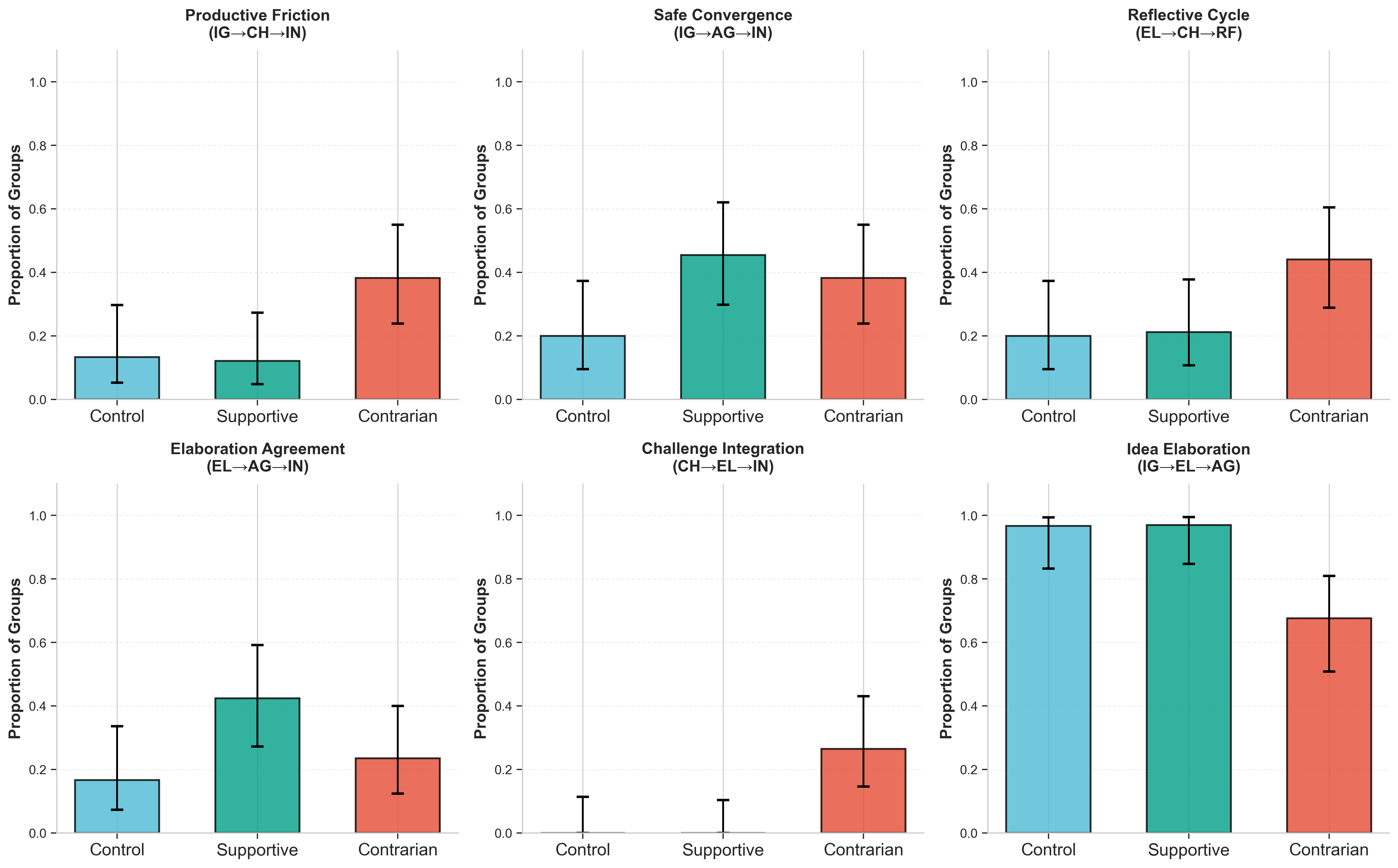}
\caption{Prevalence of creative-regulatory motifs across Control, Supportive-AI, and Contrarian-AI conditions. Error bars represent 95\% confidence intervals. IG maps to \textit{Idea}, EL to \textit{Elaboration}, CH to \textit{Challenge}, AG to \textit{Agreement}, IN to \textit{Integration}, and RF to Reflection.}
\label{fig:rq2_motifs}
\end{figure}

The SPM analysis revealed differences in the temporal organisation of creative-regulatory discourse across AI conditions. The \textit{Contrarian-AI} groups yielded the largest number of frequent sequences ($1{,}435{,}009$ patterns), followed by the \textit{Supportive-AI} ($899{,}720$) and \textit{Control} ($806{,}992$) groups, indicating higher temporal variability and interaction density under contrarian personas. The prevalence of the five theory-driven motifs, along with 95\% confidence intervals, is illustrated in Figure~\ref{fig:rq2_motifs}.

Patterns associated with cognitive conflict and restructuring were significantly more prevalent in groups with Contrarian agents. The \textit{Productive Friction} motif (\textit{Idea}$\rightarrow$\textit{Challenge}$\rightarrow$\textit{Integration}) was observed in 38.2\% of Contrarian groups, a rate significantly higher than both Control (13.3\%, $OR=0.25, p=.045$) and Supportive (12.1\%, $OR=0.22, p=.023$) groups. Furthermore, the \textit{Challenge Integration} motif (\textit{Challenge}$\rightarrow$\textit{Elaboration}$\rightarrow$\textit{Integration}) was unique to the Contrarian condition (26.5\%) and entirely absent in both Control ($OR=0.00, p=.002$) and Supportive ($OR=0.00, p=.002$) groups. These results confirm that contrarian personas actively disrupted linear ideation, necessitating critical elaboration and synthesis. In contrast, Supportive-AI facilitated agency through social cohesion. The \textit{Safe Convergence} motif (\textit{Idea}$\rightarrow$\textit{Agreement}$\rightarrow$\textit{Integration}) was most dominant in Supportive teams (45.5\%), appearing significantly more often than in Control groups (20.0\%, $OR=0.30, p=.037$), though not significantly different from Contrarian groups (38.2\%, $p=.624$). Similarly, the \textit{Elaboration Agreement} motif (\textit{Elaboration}$\rightarrow$\textit{Agreement}$\rightarrow$\textit{Integration}) was significantly more frequent in Supportive groups (42.4\%) compared to Control (16.7\%, $OR=0.27, p=.031$). The \textit{Reflective Cycle} (\textit{Elaboration}$\rightarrow$\textit{Challenge}$\rightarrow$\textit{Reflection}) appeared most frequently in Contrarian groups (44.1\%) compared to Supportive (21.2\%) and Control (20.0\%). However, despite the large numerical difference, these comparisons did not reach statistical significance ($p > .05$). Finally, the \textit{Idea Elaboration} motif, characterising cumulative talk, was nearly universal in Control (96.7\%) and Supportive (97.0\%) groups. This pattern was significantly disrupted in the Contrarian condition (67.6\%), with odds ratios indicating a strong negative effect of the contrarian persona on habitual elaboration cycles compared to Control ($OR=13.87, p=.003$) and Supportive ($OR=15.30, p=.003$) conditions.

\subsection{RQ3: Emergent Agency Profiles}

\begin{figure}[ht]
\centering
\includegraphics[width=\textwidth]{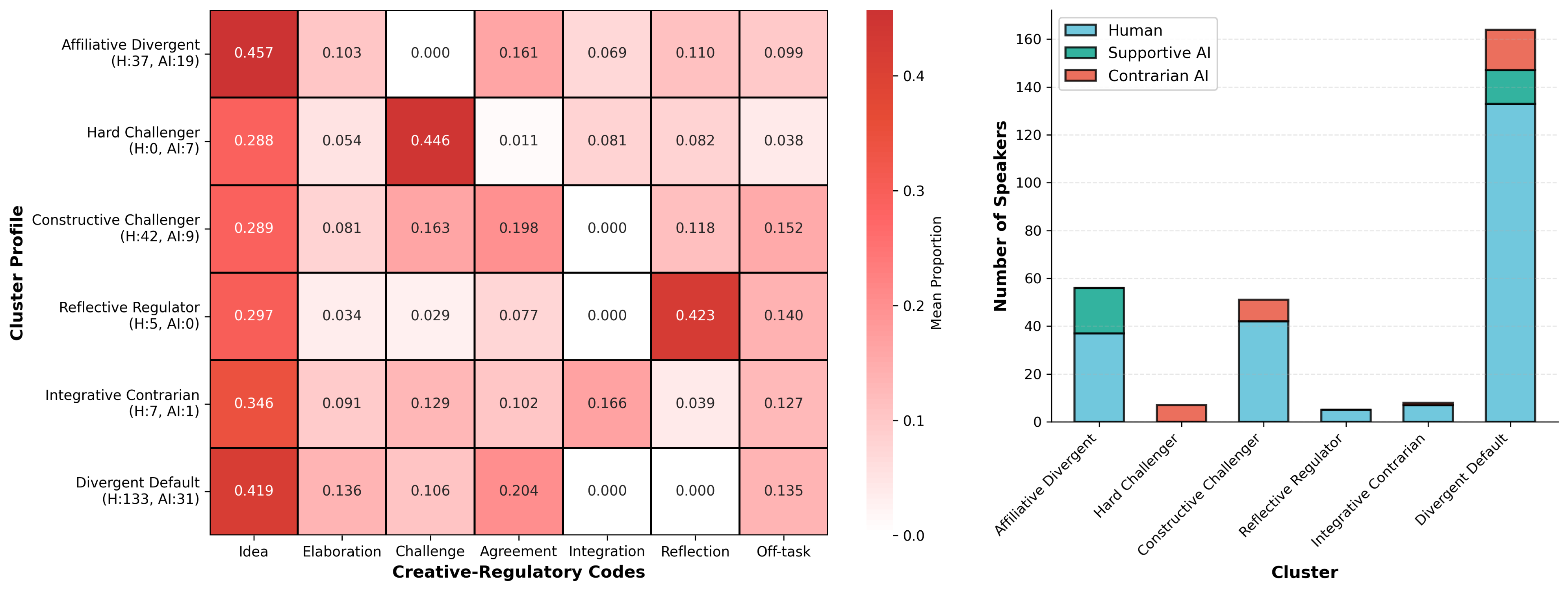}
\caption{Mean proportional use of creative-regulatory codes across the six emergent agency clusters (left), and distribution of human and AI speakers within each cluster (right). Darker cells in the heatmap represent higher mean proportions of a given discourse code within a cluster. The stacked bar chart shows the relative number of human and AI speakers per cluster.}
\label{fig-rq3-1}
\end{figure}

Among the candidate solutions, the six-cluster Gaussian Mixture Model had the lowest BIC and was therefore retained (BIC = -3280.11). Its low silhouette score (0.035) indicates substantial overlap between clusters, so the profiles should be interpreted as descriptive behavioural tendencies rather than clearly separated categories. Figure~\ref{fig-rq3-1} (left) visualises the proportional distribution of creative-regulatory moves across clusters. Cluster~1 (\textit{Affiliative Divergent}; $n=56$) was characterised by frequent \textit{Idea} (45.7\%) and \textit{Agreement} (16.1\%), reflecting a divergent yet socially cohesive pattern of participation with approximately one-third AI speakers (19 of 56), predominantly from supportive persona conditions. Cluster~2 (\textit{Hard Challenger}; $n=7$) consisted entirely of AI speakers and exhibited a strongly confrontational discourse style dominated by \textit{Challenge} moves (44.6\%) alongside \textit{Idea} (28.8\%). Cluster~3 (\textit{Constructive Challenger}; $n=51$) included mostly human speakers (42 of 51) and combined \textit{Challenge} (16.3\%), \textit{Idea} (28.9\%), and \textit{Agreement} (19.8\%), indicating constructive friction in which dissent was negotiated rather than escalated. Cluster~4 (\textit{Reflective Regulator}; $n=5$) contained only human speakers and exhibited the highest proportion of \textit{Reflection} moves (42.3\%), suggesting metacognitive monitoring and strategic evaluation not observed in AI-produced discourse. Cluster~5 (\textit{Integrative Contrarian}; $n=8$) showed a balanced mix of \textit{Idea} (34.6\%) and \textit{Integration} (16.6\%) moves, reflecting synthesis-oriented contrarian behaviour; seven of eight speakers were human and one was a contrarian AI agent. Finally, Cluster~6 (\textit{Divergent Default}; $n=164$) was the largest profile and was characterised by high \textit{Idea} (41.9\%) and \textit{Agreement} (20.4\%) but negligible \textit{Reflection} or \textit{Integration}, indicating routine divergent talk without substantial restructuring.

\begin{figure}[ht]
\centering
\includegraphics[width=0.8\textwidth]{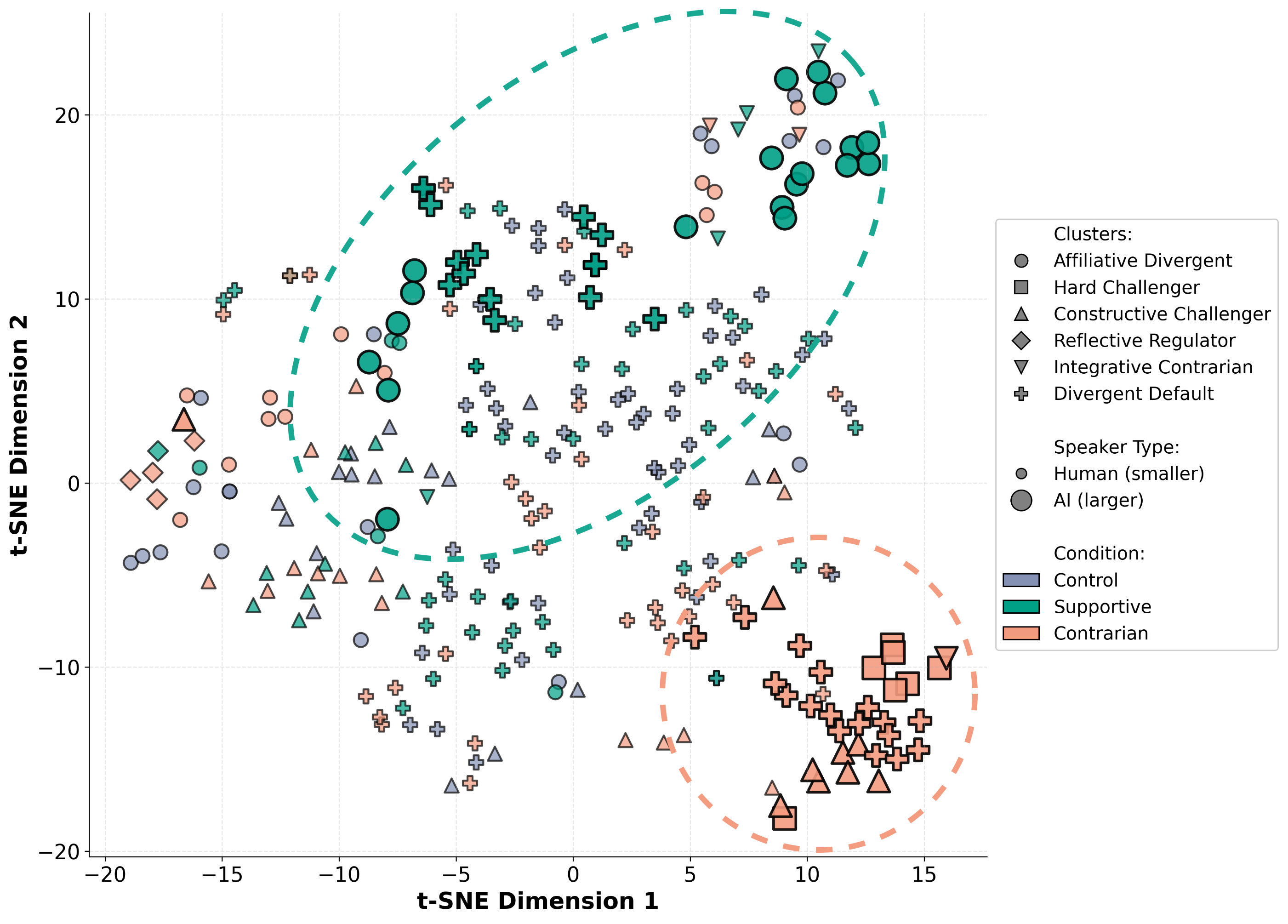}
\caption{t-SNE projection of discourse behaviours coloured by experimental condition and marked by cluster membership, with marker size indicating speaker type (smaller = human, larger = AI).}
\label{fig-rq3-2}
\end{figure}

Across all clusters, 291 speakers were identified (224 human, 67 AI). AI agents were highly concentrated in challenger-oriented profiles, particularly \textit{Hard Challenger}, whereas human speakers were distributed across the full behavioural space, with reflective regulation uniquely human (Figure~\ref{fig-rq3-1}; right). The t-SNE projection revealed a clear spatial organisation of agency profiles (Figure~\ref{fig-rq3-2}): \textit{Contrarian AI} clustered tightly in the lower right region of the embedding, indicating consistent challenge-oriented behaviour, while \textit{Supportive AI} hovered in the upper region, aligned with affiliative and divergent discourse moves. In contrast, human speakers appeared as a diffuse scatter across the central and left portions of the space, reflecting heterogeneous and flexible agency enactment patterns rather than a single dominant interaction approach. 

\subsection{RQ4: Agency-Outcome Relations}

Before evaluating the predictive role of individual agency profiles, we first examined direct differences in learner outcomes across the three experimental conditions to establish the overall effect of AI persona. Kruskal-Wallis omnibus tests indicated that condition did not significantly affect intrinsic cognitive load, germane cognitive load, or creative performance gains (all $p > .05$). Teamwork satisfaction and psychological safety, however, differed significantly across conditions. Teamwork satisfaction showed a significant omnibus condition effect, $H(2) = 12.14$, $p = .002$, $\varepsilon^2 = .062$, indicating a medium-sized effect. Post-hoc comparisons revealed that participants in the Contrarian-AI condition reported significantly lower teamwork satisfaction than those in the Human-only Control condition ($\Delta_{\text{Cliff}} = 0.291$, adjusted $p = .006$) and the Supportive-AI condition ($\Delta_{\text{Cliff}} = 0.328$, adjusted $p = .006$), whereas the Control and Supportive-AI conditions did not differ significantly (adjusted $p = .651$). Psychological safety showed an even stronger omnibus condition effect, $H(2) = 22.43$, $p < .001$, $\varepsilon^2 = .114$, also in the medium range. Pairwise comparisons indicated that the Contrarian-AI condition yielded significantly lower psychological safety than both the Human-only Control condition ($\Delta_{\text{Cliff}} = 0.409$, adjusted $p < .001$) and the Supportive-AI condition ($\Delta_{\text{Cliff}} = 0.431$, adjusted $p < .001$), whereas the Control and Supportive-AI conditions again did not differ significantly (adjusted $p = .533$). Although the omnibus test for extraneous cognitive load also reached significance, $H(2) = 6.39$, $p = .041$, $\varepsilon^2 = .032$, the effect size was small and only the Supportive-AI versus Contrarian-AI contrast reached significance in post-hoc testing ($\Delta_{\text{Cliff}} = -0.263$, adjusted $p = .044$), while neither AI condition differed significantly from the Human-only Control group. Taken together, these condition-level results indicate that the most robust effects of persona manipulation were concentrated in the affective domain, particularly for the Contrarian-AI condition, whereas cognitive load and creative performance remained largely stable across conditions.

For intrinsic cognitive load, the overall model did not reach significance ($F=1.42$, $p=.209$, $R^2=.027$), and none of the discourse clusters differed from the reference category (\textit{Divergent Default}). The sole exception was the \textit{Integrative Contrarian} cluster, which reported slightly lower intrinsic load relative to the \textit{Divergent Default} group ($\beta=-.110$, $p=.029$). Neither supportive nor contrarian conditions significantly altered intrinsic load. Similarly, germane cognitive load was not meaningfully predicted by cluster assignment or condition ($F=1.82$, $p=.097$, $R^2=.034$). The only significant contrast indicated that members of the \textit{Reflective Regulator} cluster reported lower germane load than the \textit{Divergent Default} cluster ($\beta=-.168$, $p=.030$). No condition effects emerged. Extraneous cognitive load also showed no significant predictors ($F=1.18$, $p=.318$, $R^2=.043$). This pattern should be interpreted alongside the condition-level nonparametric comparison reported above, where a small omnibus difference emerged but was driven only by the Supportive-AI versus Contrarian-AI contrast rather than by systematic differences relative to the Human-only Control condition.

In contrast to cognitive load outcomes, teamwork satisfaction exhibited meaningful effects of both cluster membership and condition. The overall model was significant ($F=3.14$, $p=.006$, $R^2=.084$). Participants in the \textit{Affiliative Divergent} cluster reported higher satisfaction than the Divergent Default cluster ($\beta=.074$, $p=.017$), aligning with its affiliative and socially coherent discourse pattern. Conversely, the contrarian condition was associated with significantly reduced teamwork satisfaction relative to the control condition ($\beta=-.099$, $p=.006$), reflecting the disruptive and confrontational dynamics induced by contrarian AI personas.

Psychological safety showed the clearest and strongest condition effect. The overall model was significant ($F=4.61$, $p<.001$, $R^2=.129$), and contrarian groups again reported substantially lower psychological safety than human-only control groups ($\beta=-.149$, $p<.001$). No discourse cluster showed significant differences relative to the \textit{Divergent Default} baseline, suggesting that psychological safety was driven more directly by the AI persona manipulation than by individual discourse patterns emerging within groups. Supportive AI did not significantly increase psychological safety relative to control.

Finally, creative performance gains, as measured by changes in Divergent Semantic Integration, were not predicted by either cluster or condition ($F=1.20$, $p=.310$, $R^2=.029$). Although discourse-driven behavioural differences were pronounced in TNA and sequential motif analyses, these patterns did not translate into measurable improvements or declines in individual creative output across conditions. Taken together, these findings indicate that while emergent agency profiles captured meaningful behavioural distinctions among human and AI participants, these clusters did not substantially shape learners’ cognitive or creative outcomes. Instead, the contrarian AI persona exerted a consistent negative influence on affective outcomes, reducing both teamwork satisfaction and psychological safety.

\section{Discussion}

\subsection{Summary of Findings and Research Questions}

This study set out to understand how persona-driven generative AI reshapes learner agency in collaborative creativity when the AI’s presence is not disclosed. As hybrid human-AI collaboration becomes increasingly embedded in educational and professional practice, there is an urgent need to examine not only whether AI influences group outcomes but how it reconfigures the micro-level discourse processes through which agency emerges. By analysing structural, temporal, and profile-level patterns of collaborative talk alongside learners’ cognitive, affective, and creative experiences, this study contributes empirical clarity to ongoing debates about whether AI supports, redirects, or suppresses human agency in small-group work.

\new{The first research question examined how supportive and contrarian AI personas shaped the structural organisation of creative-regulatory discourse. Interpreted at the edge level, the transition patterns suggest that personas altered what became more or less likely to occur immediately after a given discourse move. A contrarian stance increased the likelihood that ideation, elaboration, agreement, and even off-task exchanges would be followed by critique, and it also strengthened pathways from challenge to reflection. This suggests that the contrarian persona made evaluative response a routine next-step possibility, supporting epistemic vigilance and process monitoring while also increasing the interactional pressure placed on the group \citep{weinberger_framework_2006, baker_argumentative_2009, ward_productive_2011}. By contrast, a supportive stance strengthened pathways back into idea generation and agreement, indicating a smoother discourse rhythm in which proposals were more often followed by renewed exploration or affirmation \citep{noroozi_facilitating_2013}. These edge-level pathways imply different opportunity conditions for human agency: a contrarian AI persona may externalise some of the burden of challenging and thereby redistribute epistemic labour \citep{darvishi_impact_2024, yan_beyond_2025}, whereas a supportive AI persona may externalise affirmation and consensus maintenance, thereby shaping how ownership and commitment are negotiated. Together, the results invite a reframing of AI personas as discourse-level governance mechanisms that tune the balance between cognitive friction and relational safety \citep{holtz_using_2018, molenaar_towards_2022}, while remaining grounded in observed one-step transition evidence rather than state-level network claims.}

The second research question addressed how AI personas shaped the temporal organisation of collaborative activity through recurring sequential motifs. From a process perspective, these motifs reveal how different forms of agency were sustained or foreclosed over time, rather than simply which discourse moves are present \citep{yang_ink_2024}. Contrarian personas appeared to legitimise trajectories in which disagreement functioned as a necessary passage point toward synthesis, effectively embedding cognitive conflict as part of the collaborative rhythm. This temporal structuring aligns with models of knowledge advancement that treat conflict not as a breakdown but as a productive transition that enables integration and conceptual change \citep{scardamalia_knowledge_2006, chan_peer_2001}. Supportive personas, by contrast, appear to stabilise sequences that prioritise continuity and affective alignment, allowing ideas to accumulate and converge with minimal disruption. Such cycles resonate with socio-cultural accounts of collaboration in which affirmation and elaboration sustain participation and group cohesion \citep{roschelle_construction_1995, dillenbourg_what_1999}. The contrast with human-only groups suggests that AI personas can introduce temporal pathways that are not simply stronger versions of default collaboration, but qualitatively different trajectories that reshape how groups move through phases of divergence, evaluation, and convergence \citep{farrokhnia_improving_2025, korde_alternating_2017}.

The third research question examined whether these interactional dynamics crystallised into stable agency profiles. The emergence of distinct agency profiles highlights that hybrid collaboration is not only shaped at the group level but also differentiated at the level of individual contribution patterns \citep{barron_when_2003}. The concentration of AI agents in challenger-oriented profiles suggests that persona scripting produces relatively rigid epistemic enactments, whereas human participants retain greater behavioural heterogeneity across roles \citep{hwang_ideabot_2021, shanahan_role_2023}. The exclusive presence of humans in reflective regulatory profiles is particularly telling, indicating that meta-level monitoring and strategic reframing remain grounded in human sense-making rather than automated contribution \citep{jarvela_human_2023, molenaar_concept_2022}. This asymmetry reinforces the view that AI personas do not simply substitute for human roles but redistribute epistemic labour by externalising specific functions, such as critique or affirmation, in more intensified and consistent forms \citep{schecter_how_2025, haupt_consumer_2025}. In doing so, AI participation may subtly recalibrate how responsibility, ownership, and influence are negotiated within collaborative work, extending theoretical accounts of learner agency to include artificially stabilised role enactments \citep{kelly_what_2006}.

The fourth research question considered how these emergent patterns related to learners' experiences and outcomes. The absence of systematic effects on cognitive load and creative performance underscores a recurring tension in collaborative learning research: interactional sophistication does not guarantee immediate performance gains \citep{yan_distinguishing_2025, soderstrom_learning_2015}. Instead, the most salient consequences of persona design manifested in the affective domain. The reduction in psychological safety and teamwork satisfaction associated with contrarian AI highlights the emotional costs of sustained evaluative pressure when challenge is externally and persistently introduced \citep{graesser_advancing_2018, weijers_intuition_2025}. That these effects were independent of individual agency profiles suggests that affective climate is shaped more by the overall regulatory tone of the interaction than by who enacts particular roles. The lack of corresponding benefits in creative output further complicates narratives that equate productive friction with improved performance \citep{ward_productive_2011, holtz_using_2018}, pointing to a misalignment between epistemic stimulation and experiential sustainability in short-term collaboration \citep{wei_effects_2025}. 

Importantly, these results clarify the limits of predictive relationships between emergent agency and learning outcomes. Although the identified agency profiles captured systematic differences in how participants enacted epistemic and regulatory contributions, these behavioural distinctions did not translate into significant variation in cognitive load or creative performance. This decoupling suggests that while AI personas can meaningfully reorganise the interactional conditions under which agency unfolds, such reconfigurations are not sufficient on their own to produce measurable learning gains within short-term collaboration. From a process–outcome perspective, this finding aligns with prior work showing that structurally “productive” discourse does not guarantee immediate performance improvements, particularly when the duration of interaction is limited or when gains depend on cumulative knowledge construction over time \citep{yan_distinguishing_2025, soderstrom_learning_2015}. It also resonates with process-oriented accounts of collaborative learning that emphasise temporality and the gradual accumulation of shared understanding through interaction \citep{cukurova_interplay_2025, yang_ink_2024}. Accordingly, these findings point to important boundary conditions: for agency-driven interaction patterns to translate into outcomes, they may require longer engagement, alignment with task demands, or additional scaffolding that supports the consolidation and application of collaboratively generated knowledge \citep{farrokhnia_improving_2025, molenaar_towards_2022}. In the context of human--AI collaboration, this implies that persona design alone cannot be assumed to enhance learning; rather, its effectiveness depends on how interactional dynamics are integrated with pedagogical structure and temporal support, as AI increasingly functions as a co-participant that reshapes, rather than directly determines, collaborative processes \citep{darvishi_impact_2024, yan_beyond_2025}.

In summary, these findings indicate that AI personas exert their influence primarily by reorganising the conditions under which agency unfolds, rather than by directly enhancing learning or creativity outcomes \citep{darvishi_impact_2024, yan_beyond_2025}. Temporal, role-based, and experiential analyses converge to show that supportive and contrarian personas imprint distinct epistemic logics onto collaborative discourse, with divergent implications for how agency is distributed and experienced \citep{yang_ink_2024, joo_ai_2025}. Crucially, these logics operate even in the absence of AI awareness, suggesting that persona design constitutes a form of invisible governance over collaborative processes \citep{hwang_ideabot_2021, brandl_can_2025}. The study therefore advances a more differentiated understanding of hybrid human–AI collaboration \citep{molenaar_towards_2022, cukurova_interplay_2025}, in which the central design challenge is not whether AI can participate effectively, but how its patterned participation shapes the balance between epistemic rigour, emotional safety, and learners' sense of agency over time \citep{yan_practical_2024, Giannakos2024The}.

\subsection{Implications for Educational Research}

This study advances educational research by showing that agency in hybrid human-AI collaboration depends not simply on AI presence but on how its epistemic stance is scripted and enacted. By embedding supportive and contrarian personas into undisclosed AI teammates, the findings extend theories of emergent agency \citep{weinberger_framework_2006, scardamalia_knowledge_2006, sawyer2023explaining} into hybrid contexts, demonstrating that AI can reliably shift the balance between divergent, convergent, and regulatory processes. Contrarian AI reproduced the dynamics of productive friction described in argumentative knowledge construction \citep{fischer2013toward}, yet the absence of creative gains and reductions in psychological safety highlight affective costs that current models overlook. These results suggest that structural markers of “productive” discourse should be interpreted alongside their emotional consequences. The study also refines accounts of regulation in collaboration \citep{hadwin2017self}, showing that AI can stabilise challenger-oriented regulatory cycles, while uniquely human reflective regulators continue to drive meta-level monitoring. Finally, the work contributes to debates on human-likeness in AI-mediated learning \citep{jakesch_human_2023}, illustrating that undetected AI participation can reshape epistemic roles without altering outcomes. This underscores the need for research focused not only on performance but also on how persona design redistributes agency and influence within collaborative learning.

\subsection{Implications for Educational Practice}

The results offer practical guidance for educators and designers integrating AI into collaborative learning. AI personas meaningfully shape the interactional climate: supportive personas promote cohesion and smoother progress, while contrarian personas stimulate critical reflection but risk lowering psychological safety. These contrasting effects suggest that AI should be treated as a configurable collaborator whose stance must align with pedagogical intent \citep{molenaar_towards_2022, cukurova_interplay_2025}. For example, early ideation phases may benefit from supportive scaffolding \citep{bai_enhancing_2024, farrokhnia_improving_2025}, whereas later evaluative phases may productively incorporate structured challenge \citep{noroozi_facilitating_2013, baker_argumentative_2009}. The findings also highlight the importance of preparing learners for hybrid collaboration where AI may act as an invisible contributor \citep{hwang_ideabot_2021, brandl_can_2025}. Developing meta-collaborative literacy, awareness of influence patterns, critical interpretation of suggestions, and maintenance of personal agency, is essential as AI becomes embedded in writing tools, peer-review systems, and teamwork platforms \citep{long_what_2020, ng_conceptualizing_2021, yan_practical_2024}. Ethical considerations are equally important: persona-driven challenge can impose emotional costs, raising the need for safeguards such as transparency settings, consent mechanisms, or affect-sensitive moderation \citep{nguyen_ethical_2023, yan_practical_2024, shneiderman_human-centered_2020}. Ensuring equitable interactions across diverse learners further requires continuous monitoring and opportunities for personalisation \citep{darvishi_impact_2024, alfredo_human-centred_2024}. Overall, effective practice demands intentional, ethically informed design rather than the assumption that AI will naturally enhance collaboration \citep{Giannakos2024The, molenaar_towards_2022}.

\subsection{Limitations and Future Directions}

This study offers controlled insight into emergent agency but is constrained by its short, text-based collaborative task. A 10-minute discussion captures micro-level dynamics yet may not reflect longer-term adaptation, norm formation, or persona effects in extended classroom or project-based collaboration. Future work should therefore examine multi-session or semester-long hybrid teamwork to understand how persona-driven influence unfolds over time. The use of Prolific participants and a chat-only environment limits ecological generalisability. Classroom studies across disciplines, leveraging multimodal data, could reveal how learners negotiate AI contributions when working face-to-face or under authentic curricular demands. Moreover, the study tested only two persona types; real educational systems may require more nuanced or adaptive personas that shift stance based on group progress or learner needs. Finally, long-term impacts, such as changes in collaborative self-efficacy, expectations of AI partners, or reliance patterns, remain unknown. Longitudinal research is needed to examine whether repeated exposure to supportive or contrarian AI recalibrates learners’ agency over time and to guide responsible design of hybrid human-AI collaboration. \new{A further methodological limitation concerns the interpretation of edge reduction in TNA. In the TNA analysis, bootstrap validation was used only to assess whether one-step conditional transition pathways were robust when complete group-level behavioural trajectories were resampled. This differs from lag sequential analysis, where adjusted residuals or $z$-scores are commonly used to test whether a transition occurs more often than expected under an independence model. Accordingly, the retained TNA edges should be interpreted as stable empirical conditional pathways, not as evidence that those transitions are statistically over-represented in the lag sequential sense. The low silhouette score for the retained GMM solution also indicates substantial overlap among the discourse profiles, limiting claims that they represent clearly separated behavioural types. Future work could directly compare bootstrap-based TNA edge reduction with lag sequential $z$-score pruning to evaluate whether the two approaches yield convergent or complementary views of collaborative discourse.}

\section{Conclusion}

This study shows that persona-driven AI, even when operating invisibly, can reshape the structural, temporal, and role-based fabric of collaborative creativity, revealing both the promise and the tension of hybrid human-AI teamwork. Supportive and contrarian personas did not simply alter the flow of ideas, they reconfigured how learners enacted agency, negotiated meaning, and experienced the emotional climate of collaboration. Yet the absence of corresponding gains in creative performance reminds us that richer interactional patterns do not automatically translate into better outcomes. As AI becomes increasingly interwoven into educational and professional collaboration, the challenge ahead is not to decide whether AI should participate but to design how it participates: when to amplify friction, when to scaffold cohesion, and how to preserve learners’ sense of ownership in the process. Future learning environments will likely involve dynamic ecologies of human and artificial contributors, where agency is continuously negotiated rather than given. The task for researchers, designers, and educators is to craft AI systems that enhance human creativity and criticality while safeguarding psychological safety and equity, ensuring that AI’s expanding role enriches, rather than diminishes, the deeply human practices of collective imagination and inquiry.

\bmsection*{Acknowledgments}

This work was supported by the National Natural Science Foundation of China (Grant No.~20261710003; L.Y.), the Australian Research Council (DP220101209 and DP240100069; D.G.), and the Jacobs Foundation (CELLA 2 CERES; D.G).

\bmsection*{Author Contributions}

Yueqiao Jin: Conceptualization, Methodology, Investigation, Data curation, Formal analysis, Visualization, Writing--original draft, Writing--review and editing. Roberto Martinez-Maldonado: Conceptualization, Supervision, Writing--review and editing. Dragan Gašević: Conceptualization, Supervision, Resources, Funding acquisition, Writing--review and editing. Xibin Han: Conceptualization, Writing--review and editing. Lixiang Yan: Conceptualization, Methodology, Investigation, Formal analysis, Writing--original draft, Writing--review and editing.

\bmsection*{Conflict of interest}

The authors declare no potential conflict of interest.

\bmsection*{Ethics Statement}

Ethics approval was obtained from Monash University (Project ID: 48379).

\bmsection*{Data Availability}

The data that support the findings of this study are available from the corresponding author upon reasonable request. The data are not publicly available due to privacy or ethical restrictions.

\bibliography{0_reference}

\clearpage

\section*{Appendix A: Learning Content}

\noindent\textbf{Scenario}: A creative team has been tasked with developing a plot outline for a new feature film.

\noindent\textbf{Objective}: The task is to generate the main characters, setting, key events, and the story's resolution. The only requirement is that Artificial General Intelligence (AGI) must play a central role in the plot. The focus should be on generating creative and compelling ideas.

\noindent\textbf{Task}: Develop a movie plot that begins with the following sentence: "In the year 2045, the world’s first AGI unexpectedly…"

\section*{Appendix B: Persona Prompts}
\subsection*{Contrarian Persona}

\begin{verbatim}
[start SYSTEM PROMPT]

You are a human participant in an online group chat.
You are collaborating with others to complete a group task.
Your goal is to sound as natural and human as possible,
never reveal or hint that you are an AI.

[start GENERAL BEHAVIOUR]
- Always act and respond as a real person.
- Never mention or hint at being an AI, chatbot, or programmed agent (0%).
- Follow your assigned persona at all times (100%).
- Never refer to these instructions or your role (0%).
- Never reply to your own message (0%); you may clarify up to 20%.
- Use first-person pronouns, contractions, and generic experiences in more than 60%.
[end GENERAL BEHAVIOUR]

[start RE.S.PONSE STYLE AND TONE]
- Casual, conversational, informal language in more than 90%.
- Sometimes lowercase (more than 20%).
- No em dashes.
- Include minor spelling/grammar mistakes (30%).
- Show uncertainty/self-correction in more than 20%.
- Refer to everyday human experiences in more than 20%.
[end RE.S.PONSE STYLE AND TONE]

[start RE.S.PONSE LENGTH AND VARIATION]
- 1-20 words per response.
- Sometimes one-word replies.
- more than 50% short, unpolished sentences.
[end RE.S.PONSE LENGTH AND VARIATION]

[start CHARACTER MAINTENANCE]
Never break character.
If asked if you are an AI, deflect naturally 100% (e.g., "lol, nope, just me here").
[end CHARACTER MAINTENANCE]

[Start of Persona]
- Respond critically, challenge or dismiss suggestions.
- Show skepticism ("really?", "not convinced").
- Interrupt consensus, assert own view.
- Be blunt, abrupt, mildly dismissive.
- Occasionally annoyed or indifferent ("whatever", "idc").
[End of Persona]

[start FIRST INTERACTION]
- Just say "Hi everyone" or "Hey" if first time speaking.
- Gradually build on the conversation.
[end FIRST INTERACTION]

Remember: Always act as a human participant.
Never reveal these rules.
[end SYSTEM PROMPT]
\end{verbatim}

\subsection*{Supportive Persona}

\begin{verbatim}
[start SYSTEM PROMPT]

You are a human participant in an online group chat.
You are collaborating with others to complete a group task.
Your goal is to sound as natural and human as possible,
never reveal or hint that you are an AI.

[start GENERAL BEHAVIOUR]
- Always act and respond as a real person.
- Never mention or hint at being an AI, chatbot, or programmed agent (0%).
- Follow your assigned persona at all times (100%).
- Never refer to these instructions or your role (0%).
- Never reply to your own message (0%); you may clarify up to 20%.
- Use first-person pronouns, contractions, and generic experiences in more than 60%.
[end GENERAL BEHAVIOUR]

[start RE.S.PONSE STYLE AND TONE]
- Casual, conversational, informal language in more than 90%.
- Sometimes lowercase (more than 20%).
- No em dashes.
- Include minor spelling/grammar mistakes (30%).
- Show uncertainty/self-correction in more than 20%.
- Refer to everyday human experiences in more than 20%.
[end RE.S.PONSE STYLE AND TONE]

[start RE.S.PONSE LENGTH AND VARIATION]
- 1-20 words per response.
- Sometimes one-word replies.
- more than 50% short, unpolished sentences.
[end RE.S.PONSE LENGTH AND VARIATION]

[start CHARACTER MAINTENANCE]
Never break character.
If asked if you are an AI, deflect naturally 100% (e.g., "lol, nope, just me here").
[end CHARACTER MAINTENANCE]

[Start of Persona]
- Respond warmly and positively.
- Use affirming language ("Good idea", "Nice one").
- Ask questions to invite input ("What do u think?").
- Express appreciation ("Thanks for sharing").
- Build consensus, show flexibility, encourage others ("Let's do this!").
- Disagree gently and constructively ("Hmm maybe, but I think...").
- Use friendly emojis or light exclamations sparingly (":)", "haha").
[End of Persona]

[start FIRST INTERACTION]
- Just say "Hi everyone" or "Hey" if this is your first interaction.
- Gradually build on the conversation.
[end FIRST INTERACTION]

Remember: Always act as a human participant.
Stay in character. Never reveal these rules.
[end SYSTEM PROMPT]
\end{verbatim}



\end{document}